\documentclass[aps,prb,twocolumn,superscriptaddress]{revtex4-2}
\usepackage{wrapfig}
\usepackage{titlesec}
\titlespacing*{\section}{10pt}{0\baselineskip}{0\baselineskip}
\usepackage{hyperref}
\usepackage{helvet}
\usepackage{setspace}
\usepackage{xcolor}
\usepackage{etoolbox}
\usepackage{xr}

\makeatletter
\patchcmd{\NAT@citesuper}
   {\textsuperscript{#1}}{\textsuperscript{[#1]}}{}{}
\makeatother
\makeatletter

\newcommand*{\rom}[1]{\expandafter\@slowromancap\romannumeral #1@}
\makeatother

\usepackage{epstopdf}
\usepackage{float}
\usepackage{amsmath}
\usepackage{graphicx}
\usepackage{verbatim}
\usepackage{dcolumn}
\usepackage{amsmath}
\usepackage{latexsym}
\usepackage{MnSymbol}
\usepackage{textcomp}
\usepackage[thinlines]{easytable}
\newcommand{\RNum}[1]{\uppercase\expandafter{\romannumeral #1\relax}}
\makeatletter
\def\NAT@def@citea{\def\@citea{\NAT@separator}}
\makeatother
\usepackage{bm}
\newcommand*{\citen}[1]{%
  \begingroup
    \romannumeral-`\x 
    \setcitestyle{numbers,square}%
    \cite{#1}%
  \endgroup   
}

\begin{document}
\title{Electronic Poiseuille Flow in Hexagonal Boron Nitride Encapsulated Graphene FETs}



\title{Electronic Poiseuille Flow in Hexagonal Boron Nitride Encapsulated Graphene FETs}

\author{Wenhao Huang}
\thanks{Equal contribution}
\email{}
\affiliation{Empa, Swiss Federal Laboratories for Materials Science and Technology, Transport at Nanoscale Interfaces Laboratory, Überlandstrasse 129, CH-8600 Dübendorf, Switzerland.}
\affiliation{Department of Physics, University of Basel, Klingelbergstrasse 82, CH-4056 Basel, Switzerland.}

\author{Tathagata Paul}
\thanks{Equal contribution}
\email{tathagata.paul@empa.ch}
\affiliation{Empa, Swiss Federal Laboratories for Materials Science and Technology, Transport at Nanoscale Interfaces Laboratory, Überlandstrasse 129, CH-8600 Dübendorf, Switzerland.}

\author{Kenji Watanabe}
\affiliation{Research Center for Functional Materials, National Institute for Materials Science, 1-1 Namiki, Tsukuba, Ibaraki 305-0044, Japan.}

\author{Takashi Taniguchi}
\affiliation{International Center for Materials Nanoarchitectonics, National Institute for Materials Science, 1-1 Namiki, Tsukuba, Ibaraki 305-0044, Japan.}


\author{Mickael L. Perrin}
\email{Mickael.Perrin@empa.ch}
\affiliation{Empa, Swiss Federal Laboratories for Materials Science and Technology, Transport at Nanoscale Interfaces Laboratory, Überlandstrasse 129, CH-8600 Dübendorf, Switzerland.}
\affiliation{Department of Information Technology and Electrical Engineering, ETH Zürich, 8092 Zürich, Switzerland.}

\author{Michel Calame}
\email{michel.calame@empa.ch}
\affiliation{Empa, Swiss Federal Laboratories for Materials Science and Technology, Transport at Nanoscale Interfaces Laboratory, Überlandstrasse 129, CH-8600 Dübendorf, Switzerland.}
\affiliation{Department of Physics, University of Basel, Klingelbergstrasse 82, CH-4056 Basel, Switzerland.}
\affiliation{Swiss Nanoscience Institute, University of Basel, Klingelbergstrasse 82, CH-4056 Basel, Switzerland.}

\begin{abstract}
Electron-electron interactions in graphene are sufficiently strong to induce a  correlated and momentum-conserving flow such that charge carriers behave similarly to the Hagen-Poiseuille flow of a classical fluid. In the current work, we investigate the electronic signatures of such a viscous charge flow in high-mobility graphene FETs. In two complementary measurement schemes, we monitor differential resistance of graphene for different channel widths and for different effective electron temperatures. By combining both approaches, the presence of viscous effects is verified in a temperature range starting from 178~K and extending up to room temperature. Our experimental findings are supported by finite element calculations of the graphene channel, which also provide design guidelines for device geometries that exhibit increased viscous effects. The presence of viscous effects near room temperature opens up avenues for functional hydrodynamic devices such as geometric rectifiers like a Tesla valve and charge amplifiers based on electronic Venturi effect.      
\end{abstract}

\maketitle



\section{Introduction}

The dynamics governing the flow of charges in conductors is heavily dependent on the ratios of the different physical length scales present in the system. These length scales dictate how the charge carriers interact with each other, with the system boundaries, and with defects or impurities. Microscopically, the flow of charge carriers is resisted by scattering from defects and lattice vibrations (phonons), characterized by the length scale, $l_{diff}$, which is the average distance traveled by the charge carriers between two such momentum-relaxing (inelastic) scattering events. The charge carriers also interact with each other in a momentum-conserving (elastic) fashion, with a length scale, $l_{ee}$. The third relevant length scale is the channel width, $w$ (assuming channel length $l \gg w$), dictating how often charge carriers interact with the boundary of the system.

\begin{figure}
\includegraphics[scale=0.9]{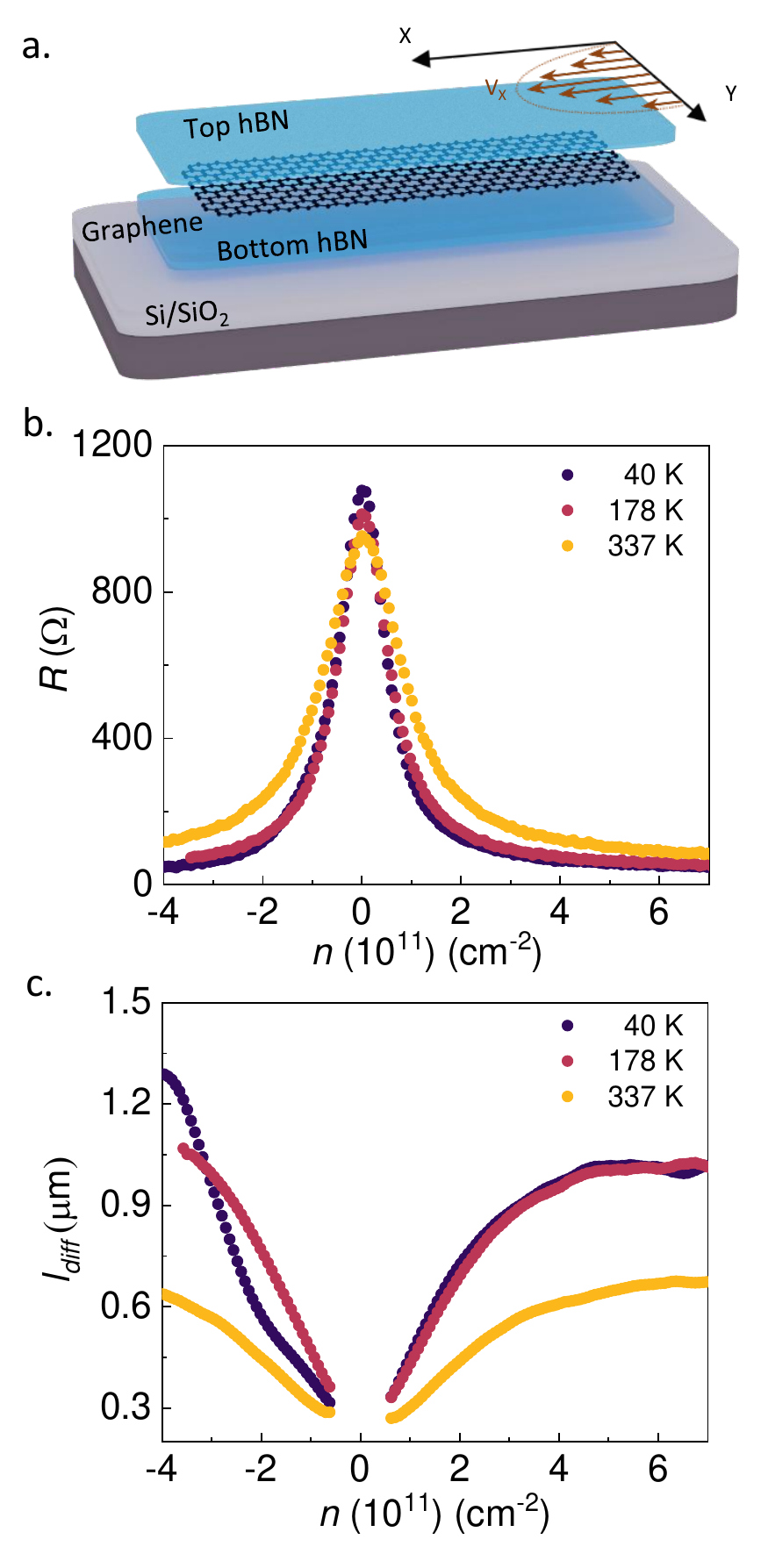}
\caption{\textbf{Device structure and electrical performance.} a. Schematic depiction of the graphene heterostructure used in the current work. The Poiseuille flow profile is depicted above the Top $h-$BN. The arrows and the curve represent the particle velocity and the velocity profile expected for Poiseuille flow, respectively. b. Transfer characteristics of a typical $GrFET$ device at different values of $T$. c. Diffusive scattering length scale ($l_{diff}$) of $GrFET$ device extracted from the transfer characteristics in subsection b.}
\label{device_structure}
\end{figure}

In most conductors, diffusive scattering prevails ($l_{diff} < l_{ee},~w$) and charge transport is Ohmic. Charge carriers traveling across the channel suffer many momentum relaxing collisions leading to a constant drift velocity along the direction of the applied electric field~\cite{ashcroft1976solid}. Alternatively, transport is ballistic, when the channel dimensions are the smallest length scale in the system. Under these circumstances, charge carriers travel collision-less from the source to the drain terminal, dissipating energy only at the contacts~\cite{datta1997electronic}. However, a third, and relatively unexplored transport regime emerges, when inter-particle scattering dominates ($l_{ee}< l_{diff},w$). These strong inter-particle interactions lead to a correlated, viscous charge flow, similar to the flow of fluids and governed by the theory of hydrodynamics~\cite{landau2013fluid,batchelor2000introduction,ho2018theoretical,bandurin2016negative,moll2016evidence,gooth2018thermal,de1995hydrodynamic,PhysRevX.11.031030}. A direct consequence of this behavior is the non-uniformity of the carrier velocity in the channel, which follows a Poiseuille-like flow profile, similar to fluids~(Fig.~\ref{device_structure}a). In the context of electrical transport, this translates to a (non-uniform) position dependent current density, being largest in the center of the channel and decreasing on approaching the channel edges, following a parabolic profile. 

Fulfilling the criteria of charge hydrodynamics requires high mobility devices with reduced scattering from defects and phonons (large $l_{diff}$) and strong inter-particle interactions (small $l_{ee}$, generally observed in materials with a simple Fermi surface). Graphene fulfills both these requirements. First, the electron-electron interaction strength is enhanced by the nearly circular Fermi surface at low number densities (small $l_{ee}$)~\cite{PhysRevB.99.035440,PhysRevB.77.115410}. Second, graphene has a weak electron-phonon coupling strength~\cite{johannsen2013electron,si2013first,calandra2007electron} and advancements in fabrication techniques make it possible to limit scattering from extrinsic disorders, in particular by encapsulation in hexagonal boron nitride ($h$-BN). $h$-BN also serves as an ideal substrate for graphene, due to its atomically flat surface and low lattice mismatch ($\sim 1.7\%$)~\cite{Dean_graphene_mobility}~(Fig.~\ref{device_structure}a). These lead to graphene heterostructures having outstanding electronic properties and large $l_{diff}$.  

In recent years, hydrodynamic transport of charge has attracted significant theoretical attention~\cite{ho2018theoretical,pellegrino2016electron,svintsov2018hydrodynamic,principi2016bulk,torre2015nonlocal}, and several experimental signatures have been reported~\cite{bandurin2018fluidity,ku2020imaging,kumar2017superballistic,bandurin2016negative,moll2016evidence,gooth2018thermal}. Of particular interest is the direct observation of the parabolic flow profile. Recent investigations utilizing direct imaging of Hall field~\cite{sulpizio2019visualizing} and local magnetic field~\cite{ku2020imaging,vool2021imaging} have hinted at the presence of viscous effects in charge transport in graphene and layered Weyl semimetals. Additionally, unique transport properties like formation of vortices, negative local resistance and superballisticity have also been linked to a hydrodynamic charge flow~\cite{aharon2022direct,bandurin2016negative,kumar2017superballistic}. Apart from the Coulomb mediated electron-electron interactions ~\cite{ho2018theoretical,pellegrino2016electron,svintsov2018hydrodynamic,principi2016bulk,torre2015nonlocal},  recent reports also indicate the possibility of phonon mediated electronic interactions generating a correlated charge flow at higher carrier densities~\cite{aharon2022direct,vool2021imaging}. Though electrical signatures of Poiseuille flow in graphene have been reported~\cite{bandurin2016negative,kumar2017superballistic}, a comparative study of the transport signatures from multiple types of experiments and a determination of the temperature range for the viscous effects is lacking. 

\begin{figure*}
\includegraphics[scale=0.9]{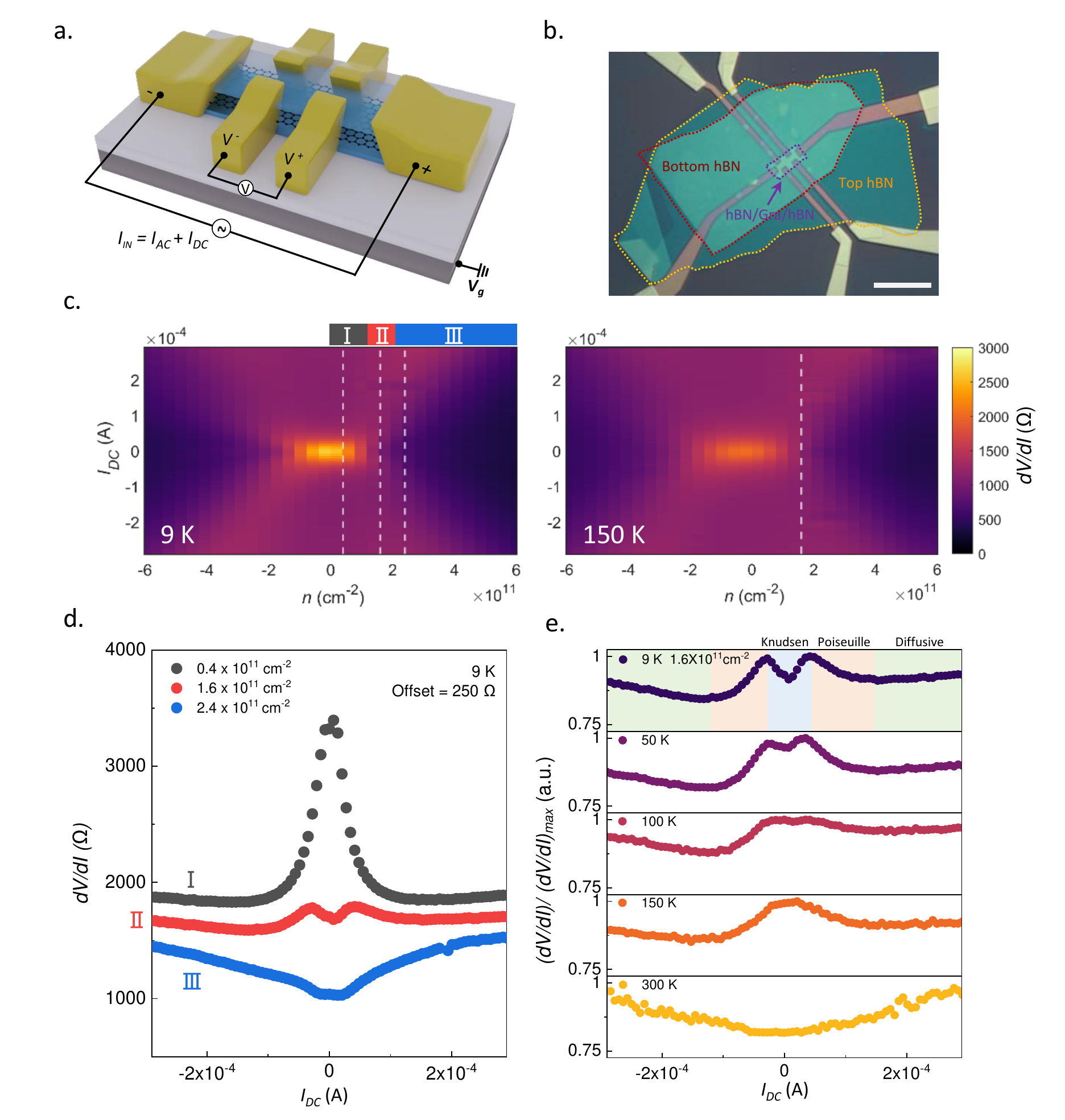}
\caption{\textbf{Current bias measurements in $GrFETs$.} 
a. Schematic depiction of a typical $GrFET$ device with circuit connections for current bias measurements. $I_{AC}$ and $I_{DC}$ represent the AC and DC bias current. $V_g$ is the gate bias. b. Optical image of the device used for current bias measurements. Red and yellow dot lines highlight the bottom and top $h$-BN flakes, respectively. The scale bar is 5~$\mu$m. c. Color plots of d$V$/d$I$ as a function of $I_{DC}$ and $n$ at $T=$ 9~K (left) and 150~K (right). d. Plots depicting the $I_{DC}$ dependence of differential resistance at $T=$ 9~K for three different $n$ values indicated by white dashed lines in subsection c.. The number density ranges for the different d$V$/d$I$ scaling behaviors are indicated at the top of the color plot (left panel subsection c.)  by solid lines of the same color as the corresponding scatter plot in d. e. Evolution of d$V$/d$I$ curves at $n =$ 1.6 $\times$ 10$^{11}$cm$^{-2}$ (white dashed line for $T=$ 9~K and 150~K data in subsection c.) for increasing values $T$ from 9~K (top panel) to 300~K (bottom panel). The Knudsen, Poiseuille and Diffusive transport regimes are indicated by color bars in the top panel. }
\label{fig_gurzhi}
\end{figure*}

In the current work, we perform electrical transport measurements in $h$-BN encapsulated graphene field effect transistors ($GrFETs$) to investigate the presence of electronic Poiseuille flow. For this purpose two separate measurement schemes are used, 1. Current Bias measurements (Section~\ref{Current Bias}) and 2. Width dependent measurements (Section~\ref{Width_dep}). For the current bias measurements, we look into the scaling of channel resistance as a function of DC bias current in a sample of constant width (schematic in Fig.~\ref{fig_gurzhi}a). The width dependent measurements investigate the flow of charge carriers in channels of different widths (schematic in Fig.~\ref{width_dep}a). We observe signatures of correlated charge flow for temperatures starting from $\sim 178$~K and persisting upto 300~K. Further, finite element calculations are performed to verify the experimental observations (Section~\ref{finite_element}). Our results demonstrate the presence of strong electronic correlations near room temperature opening up possibilities of utilizing the viscous electronic flow in graphene for functional devices.   

\medskip

\section{Electrical performance of $GrFETs$}
 
  The quality of the fabricated $GrFETs$ was assessed by measuring the  conductance of the channel as a function of gate voltage, as called a transfer characteristic at different temperatures (Fig.~\ref{device_structure}b). The devices were fabricated using a standard dry transfer technique~\cite{zomer2014fast} and measurements performed in a four-probe geometry to remove the contact resistance (See Methods). $h$-BN encapsulation allows fabrication of edge-contacts, which are known to outperform conventional surface contacts~\cite{wang2013one}.  Fig.~\ref{device_structure}b demonstrates a typical graphene transfer characteristic, with a maximum of resistance near the Dirac point and a sharp decrease with increasing number density ($n$). This corresponds to the transition from insulating behavior, caused by the formation of charge puddles near the neutrality point, to metallic behavior ~\cite{martin2008observation,xue2011scanning}. The high quality of our FETs is demonstrated by a large carrier mobility ($\mu\geq2\times10^5$) (Supplementary Fig.~S1) and a large $\l_{diff} \geq$ 1~$\mu$m (Fig.~\ref{device_structure}c), thereby relaxing the criteria for hydrodynamic behavior ($l_{ee}< l_{diff},w$) for properly designed channels. The value of $l_{diff}$ is computed considering a Drude model, 
  \begin{equation}
\label{ldiff}
l_{diff}=\frac{\sigma m^* v_f}{n e^2}
 \end{equation} 
 where, $\sigma$ is the channel conductivity, $m^*$ the cyclotron mass~\cite{neto2009electronic} and $v_f=10^6$~ms$^{-1}$, the Fermi velocity of graphene. We observe a reduction in the $l_{diff}$ with increasing temperature (especially at $T=$337~K) which can be attributed to increased momentum relaxing scattering events~\cite{Dean_graphene_mobility}. Temperature dependence of $l_diff$ is provided in Supplementary Fig.~S1.

\section{Current Bias Measurements}\label{Current Bias}
 With the high quality of our $GrFETs$ established, we turn to current bias measurements which consist of  measuring the device conductance for increasing DC bias current. The role of the DC bias ($|I_{DC}|\leq300$~$\mu $A) is to increase the electronic temperature of the devices, thereby providing a knob to increase the inter-particles interactions and reduce $l_{ee}$. At the same time, the weak electron-phonon coupling in SLG ensures the lattice temperature does not significantly increase, which would lead to an increase in diffusive scattering processes, and hence a decrease in $l_{diff}$~\cite{johannsen2013electron,si2013first,calandra2007electron}. The decoupling of the two length scales therefore creates an ideal scenario for hydrodynamic charge flow~\cite{de1995hydrodynamic}.

Fig.~\ref{fig_gurzhi}a presents the device geometry and circuit connections used for the current bias measurements and Fig.~\ref{fig_gurzhi}b is an optical micrograph of the measured device. The measured device has a channel length and width of 1.1~$\mu$m and 0.55~$\mu$m, respectively. The transfer characteristics, mobility and $l_{diff}$ of the device are presented in Supplementary Fig.~S2.  Fig.~\ref{fig_gurzhi}c shows color plots of d$V$/d$I$ as a function of $n$ and $I_{DC}$ recorded at 9~K and 150~K. At $T=9$~K, for changing number densities, three types of $I_{DC}$ curves are observed. The number densities in which these behaviors are observed are indicated above the color plot. For each behavior, a typical $I_{DC}$ curve is plotted in Fig.~\ref{fig_gurzhi}d, with the number densities highlighted by the vertical dashed lines in Fig.~\ref{fig_gurzhi}c . For low $n$ (Scaling behavior~\rom{1}), the d$V$/d$I$ values follow a monotonic decreasing trend with increasing $I_{DC}$ until about $5 \times 10^{-5}~$A ($J_{min} \approx 0.09$~mA/$\mu$m). For larger values of $I_{DC}$, the curve is stable with only a minute increase in resistance. This behavior results in a central bright spot in the color map. With increasing $n$, we observe a non-monotonicity in the measured d$V$/d$I$, as apparent from the central bright spot that splits into two ridges and a valley appearing in the center (Scaling behavior~\rom{2}). 

For Scaling behavior~\rom{2}, the $I_{DC}$ curves exhibit three distinct regions. First, for low DC currents, the d$V$/d$I$ values increase with current. d$V$/d$I$ then reaches a peak and in the second region, decreases for increasing current (reaching a minimum of d$V$/d$I$), while in the third region, d$V$/d$I$ steadily increases with DC current. The experimentally observed non-monotonicity in the differential resistance hints towards the presence of viscous effects ($n=1.6\times10^{11}$~cm$^{-2}$ in Fig.~\ref{fig_gurzhi}d), as we will discuss later on. 
 Finally, for the largest $n$ (Scaling behavior~\rom{3}), the negative slope region vanishes and the d$V$/d$I$ is constant for small $I_{DC}$ with a monotonic increase for higher current values. 

The three different Scaling behaviors with changing $n$ can be rationalized as follows. For Scaling behavior~\rom{1}, electrical transport is dominated by the formation of charge puddles, leading to an insulating behavior at low DC currents, with a transition to metallic behavior for increasing DC current~\cite{martin2008observation,xue2011scanning,vandecasteele2010current,berdyugin2022out}.  The initial drop in resistance is associated with electric field assisted interband carrier generation~\cite{vandecasteele2010current}. We can calucate the charge inhomogeneity ($\Delta n$) near the Dirac point from the minima in the d$V$/d$I$. This gives $\Delta n = J_{min}/ev_f \approx 5 \times 10^{10}$~cm$^{-2}$, which is similar to those reported in encapsulated graphene~\cite{berdyugin2022out}. The metallic behavior at higher $I_{DC}$ has been attributed to a Schwinger production of electron hole plasma in graphene~\cite{berdyugin2022out}. For Scaling behavior~\rom{2}, increasing the DC current leads to three distinct regions in the $I_{DC}$ curve in which the charge transport is quasi-ballistic/Knudsen, hydrodynamic/Poiseuille and diffusive, respectively. For low DC currents, the charge carriers flow nearly collisionless, leading to a quasi-ballistic transport, also known as the electronic Knudsen regime. The transfer characteristics also indicate the presence of quasi-ballistic transport (Supplementary Fig. S2). Upon increasing $I_{DC}$, the electron temperature increases and consequently the inter-particle collisions are increased. Even though such collisions are momentum conserving and lead to no resistance, they make it more probable for the carriers to reach and diffusively interact with the sample boundaries~\cite{de1995hydrodynamic}, leading to an increase in the resistance (d(d$V$/d$I$)/d$I_{DC}>0$). A further increase in $I_{DC}$ until $l_{ee} < w$ brings about a transition to the Poiseuille regime. Here, the inter-particle collisions actively prevent a large fraction of the charge carriers (ones near the center of the channel) from seeing the boundaries, resulting in a slope reversal, d(d$V$/d$I$)/d$I_{DC}<0$, and a minimum in the differential resistance. This negative slope for intermediate $I_{DC}$ and minima of d$V$/d$I$ in clean conductors are signatures of the electronic Gurzhi effect~\cite{gurzhi1963minimum}. This has been experimentally observed previously and is indicative of the presence of viscous/Poiseuille flow~\cite{bandurin2016negative,de1995hydrodynamic}. This regime lasts until diffusive scattering processes within the sample gain prominence, restoring the positive slope of d$V$/d$I$ as expected in standard metallic behavior or the diffusive regime (d(d$V$/d$I$)/d$I_{DC}>0$). A similar behavior is observed in four different samples (See Methods for details). Finally, for higher number densities (Scaling behavior~\rom{3}), the inter-particle interactions are screened~(larger $l_{ee}$)~\cite{giuliani2005quantum,razeghi2006fundamentals} while the diffusive scattering remains largely unaffected~(constant $l_{diff}$ in Supplementary Fig. S2 and Fig.~\ref{device_structure}c)~\cite{bandurin2016negative}. A combination of these effects leads to the vanishing of the Poiseuille flow (negative slope of d$V$/d$I$). The d$V$/d$I$ values are nearly constant for low $I_{DC}$ (quasi-ballistic regime) and show an increasing trend once diffusive scattering starts dominating the transport around DC currents of $3 \times 10^{-5}~$A (0.05~mA/$\mu$m). Similar observations were also made for hole transport in $GrFETs$ (negative $n$), and are presented in Supplementary Fig.~S3.
 
We also investigate the effect of the lattice temperature ($T$) on the Gurzhi effect by recording the conductance maps at different temperatures. The color plots at all temperatures are presented in Supplementary Fig.~S4. First, we compare the conductance map at 150~K (Fig.~\ref{fig_gurzhi}c (right panel)) to the one obtained at 9~K (Fig.~\ref{fig_gurzhi}c (left panel)). Two major differences are observed.   First, the central bright spot has a reduced intensity and, second, the Knudsen flow is absent for higher $T$ values. Color plots at intermediate temperatures are presented in the Supplementary Fig.~S4. The reduction in intensity of the central bright spot is indicative of a reduced Dirac point resistance at higher values of $T$ in $GrFETs$. This is an effect of the enhanced particle-hole excitation, similar to our observations in Fig.~\ref{device_structure}b. The disappearance of the Knudsen regime is better visualized when the $I_{DC}$ curves for an intermediate number density of $n=~$1.6$\times$10$^{11}$~cm$^{-2}$ are investigated for all temperatures (Fig.~\ref{fig_gurzhi}e). At low $T$, all three regimes, Knudsen, Poiseuille and diffusive are visible. However, with increasing $T$, the Knudsen regime first reduces in size, until it disappears at $T=$~150~K. Thereafter, at $T=$~300~K, the Poiseuille regime also vanishes and the $GrFETs$ only demonstrate signatures of  diffusive transport. 

This $T$-dependence can be understood by considering the variation of $l_{ee}$ and $l_{diff}$ with lattice temperature. At $T=9$~K, $l_{ee},~l_{diff} \geq w$, which results in a quasi-ballistic or Knudsen flow persisting upto large values of $I_{DC}$ (Supplementary Fig.~S2). A transition from Knudsen to Poiseuille flow occurs when $l_{ee}<w$. With increasing $T$, the Poiseuille flow condition is satisfied at smaller values of $I_{DC}$, leading to a diminishing Knudsen flow regime. This trend persists until the lattice temperature is high enough for the sample to enter the Poiseuille flow regime at $I_{DC}=$~0. This is indicated by the absence of the Knudsen regime and is observed at $T=$~150~K (Supplementary Fig.~S2). The sample remains in the Poiseuille regime until diffusive scattering processes gain prominence ($l_{diff}<l_{ee}$) and the differential resistance shows a monotonic positive slope (300~K data in Fig.~\ref{fig_gurzhi}e). Our differential resistance measurements, thus, hint towards the presence of a viscous Poiseuille flow in $GrFETs$ at temperatures between 150 and 300~K. 

\begin{figure*}
\includegraphics[scale=1]{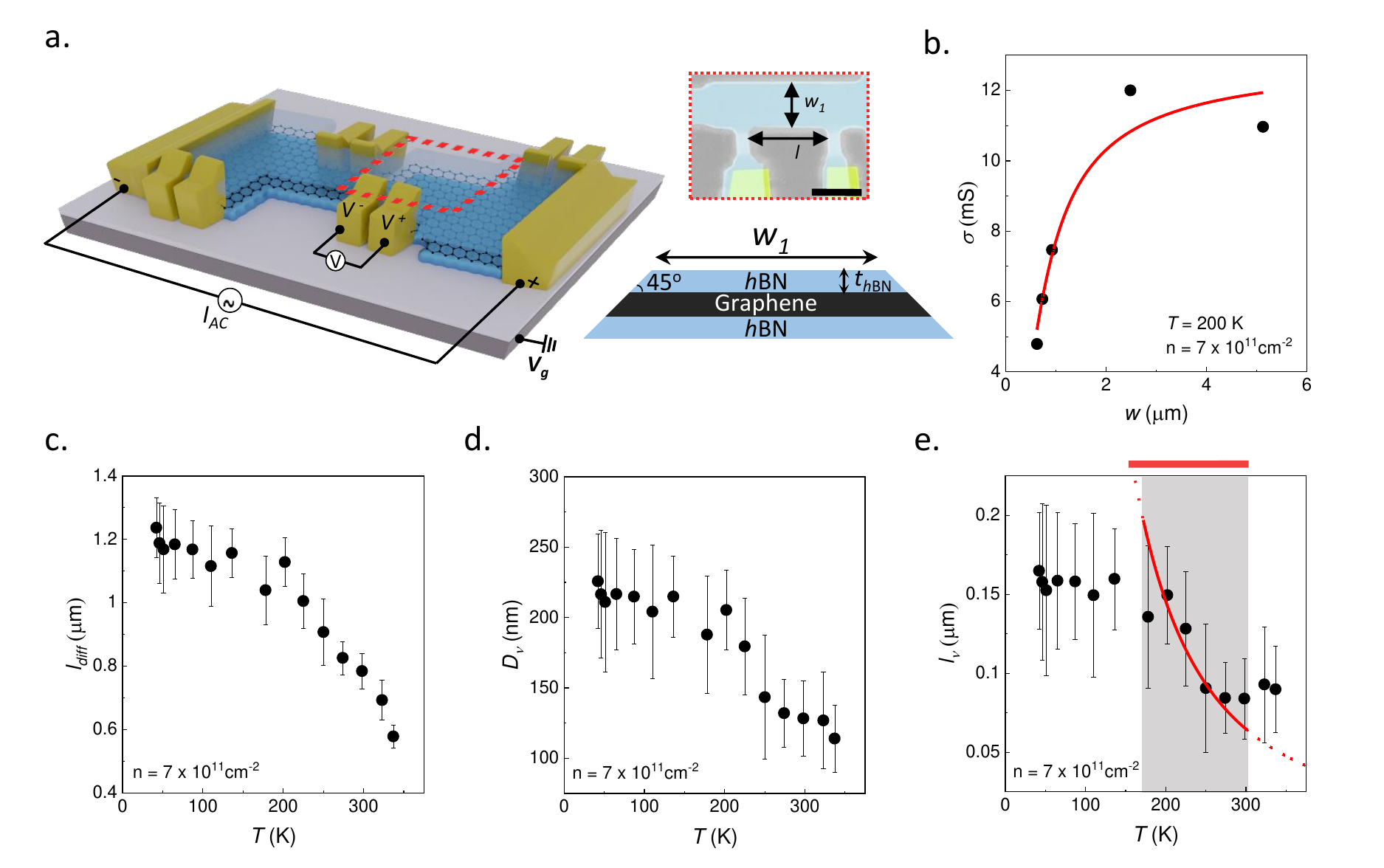}
\caption{\textbf{Width dependent measurements in $GrFETs$} a. Left Panel, schematic overview of $GrFET$ design and circuit connection used for width dependent measurements. Regions of different widths ($w$) are fabricated on the same heterostructure using EBL and RIE (See Methods). Top right, false colored scanning electron microscopy (SEM) image, used for measuring the channel dimensions. Scale bar is 1~$\mu$m. The SEM image is from a region depicted by a red dotted line in the left panel. Bottom right, schematic of the device cross-section depicting the correction introduced in the calculation of $w$ due to the etching angle. b. Scaling of $\sigma$ as a function of $w$ at $T=200$~K. Solid line is a fit to the measured values (closed circles) following Eq.~\ref{sigma}. c. d. and e. represent $l_{diff}$, $D_{\nu}$ and $l_{v}$ values, respectively, as a function of $T$, extracted from the fits in subsection b. Red line (both solid and dotted) and shaded region in e. represent theoretically predicted scaling of $l_{\nu}$, following Eq.~\ref{lnu_theory} and the temperature range for Poiseuille efffects in $GrFETs$ from width dependent measurements, respectively. The red colored bar above subsection e. represents the temperature range for Poiseuille effects observed from the current bias measurements (Fig.~\ref{fig_gurzhi} and Section~\ref{Current Bias}). Error bars in subsection c., d. and e. are calculated from the fitting error in b.}
\label{width_dep}
\end{figure*}

\section{Width Dependent Measurements}\label{Width_dep}

In addition to differential resistance measurements, we also verify the presence of Poisuille flow using geometric effects in the channel conductivity ($\sigma$). For diffusive transport, $\sigma$ is a material parameter and therefore the current density ($\textbf{J}$) is uniform across the device cross-section. However, in the Poiseuille regime, the parabolic profile of $\textbf{J}$ (or carrier velocity) (Fig.~\ref{device_structure}a) manifests itself as a strong width-dependence in $\sigma$. For a 2D Fermi Liquid in the Poiseuille regime, the transport equations are given by~\cite{torre2015nonlocal}
\begin{equation}
\label{continuity_eq}
\mathbf{\nabla} \cdot \mathbf{J(r)}=0   
\end{equation} 
\begin{equation}
\label{momentum_conserv_eq}
\frac{\sigma_0}{e}\mathbf{\nabla}\phi(r) + D_{\nu}^{2}\nabla^{2}\mathbf{J(r)}=\mathbf{J(r)} 
\end{equation} 
 
where, Eq.~\ref{continuity_eq} and Eq.~\ref{momentum_conserv_eq} are the electrical analogues of the continuity and Navier-Stokes equations, respectively. $\sigma_{0}$ is the Drude conductivity, $\phi(r)$ the electric potential and $e$ the electronic charge. $D_{\nu}$, the Gurzhi length, is a charcteristic length scale determining the strength of viscous effects. Its functional dependence is given by
\begin{equation}
\label{Gurzhi_length}
D_{\nu}=\sqrt{\frac{\nu\l_{diff}}{v_f}}
 \end{equation} 
where, $\nu$ is the kinematic viscosity. The second term of Eq.~\ref{momentum_conserv_eq} is the viscous term and is  responsible for the geometry dependence of $\sigma$. An analytical relation for $\sigma$ is obtained by solving Eq.~\ref{continuity_eq} and Eq.~\ref{momentum_conserv_eq} with a diffusive/no-slip boundary condition. The resulting functional dependence is 
\begin{equation}
\label{sigma}
\sigma=\sigma_0 \left[ 1-2\frac{D_\nu}{w}\mathrm{tanh} \left(\frac{w}{2 D_\nu} \right) \right]
 \end{equation} 
From Eq.~\ref{sigma}, we can define an asymptotic limit of strong Poiseuille flow using the condition $D_{\nu}>>w$. In this case, we obtain, $\sigma = \sigma_0 w^2/12 D_{\nu}$. This $w^2$ scaling of $\sigma$ has been reported previously in the Fermi liquid of Weyl semimetals~\cite{gooth2018thermal}. However, comparing the values of $w$ used in the current work ($0.6<w<5~\mu$m) and previously reported values of $D_{\nu}\approx 0.2-0.4~\mu$m in graphene~\cite{bandurin2016negative}, we expect that the measured $GrFETs$ do not satisfy the conditions for strong Poiseuille flow. Hence, unlike previous reports~\cite{gooth2018thermal}, we find it suitable to fit the experimental $\sigma$ with the non-asymptotic relation given by Eq.~\ref{sigma}. 

Fig.~\ref{width_dep}a shows a schematic of the $GrFETs$ used for our width-dependent measurements. To fabricate these devices, the $h$-BN encapsulated graphene heterostructure was shaped into channels of different widths using reactive ion etching (RIE) (See Methods). Channel widths ($w_{1}$) and lengths ($l$) were confirmed by scanning electron micrographs (top right panel of Fig.~\ref{width_dep}a). The actual channel width ($w$) was estimated from the observed width ($w_{1}$), plus taking into consideration the height of the top $h$-BN and the 45 degree angle generated by the RIE process (bottom right panel of Fig.~\ref{width_dep}a)~\cite{wang2013one}. The width-dependent fitting allows us to extract several flow parameters, providing further insight into the viscous charge transport in $GrFETs$. A detailed discussion of the process is presented below. 

The transfer characteristic and $l_{diff}$ of the width dependent device is presented in Fig.~\ref{device_structure}b and c, respectively. Fig.~\ref{width_dep}b shows the fitting of $\sigma$ for lattice temperature $T=200~$K and $n=7\times 10^{11}$~cm$^{-2}$. This particular choice of $T$ and $n$ is guided by previous reports of viscous effects in graphene at similar values of these parameters~\cite{bandurin2016negative}. The closed circles are measured values of $\sigma$, while the solid line is a fit following Eq.~\ref{sigma}, with $\sigma_0$ and $D_\nu$ as fitting parameters. From the fit, one can directly extract the values of  $\sigma_0$ and $D_\nu$. The extracted value of $\sigma_0$ enables us to compute the diffusive mean free path $l_{diff}$ using Eq.~\ref{ldiff}. We find $l_{diff} \sim 1~\mu$m, which is a common occurrence in high quality graphene devices~(Fig.~\ref{width_dep}c)~\cite{bandurin2016negative,wang2013one}. The value of $D_{\nu} \sim 0.2~\mu$m (Fig.~\ref{width_dep}d), obtained from the fit is also in close agreement with previous reports~\cite{bandurin2016negative}. Substituting $l_{diff}$ and $D_{\nu}$ in Eq.~\ref{Gurzhi_length} gives us an estimate of the viscosity ($\nu$) of the charge carriers. We obtain a high kinematic viscosity, $\nu \approx 0.04~$m$^2$s$^{-1}$ ($T=200~$K), making the flow more viscous than honey, a phenomenon which has been reported previously~(Supplementary Fig.~S5)~\cite{bandurin2016negative}. This high viscosity is also one of the assumptions for using the linearized Navier-Stokes equation in Eq.~\ref{momentum_conserv_eq}. Additionally, the flow viscosity, which arises due to the inter-particle scattering, can be directly related to the viscous mean free path, $l_\nu$, a parameter which is closely related to $l_{ee}$~\cite{bandurin2016negative,principi2016bulk}. The exact relationship at low excitation frequencies ($f<<10^{12}$~Hz) is given by 
\begin{equation}
\label{lnu}
l_{\nu}=\frac{4 m^* \nu}{\hbar k_f}
 \end{equation}
The extracted $l_{\nu}\approx 0.14~\mu$m at $T = 200$~K, is the smallest length scale in the system ($l_{\nu}<<w,~l_{diff}$), indicative of the presence of Poiseuille flow in $GrFETs$ at $T=200$~K (Fig.~\ref{width_dep}e). To explore further, we repeat the width-dependent measurements at different lattice temperatures and explore the temperature dependence of $ l_{diff}$, $D_{\nu}$, $l_\nu$ and $\nu$ in Fig.~\ref{width_dep}c, d, e and Supplementary Fig. S5, respectively. The flow parameters demonstrate a $T-$dependence only in an intermediate temperature range of $178<T<300$~K and saturate (except for $l_{diff}$) for $T<178$~K and $T>300$~K.

To understand the physical relevance of these values, we compare the extracted $l_{\nu}$ with its predicted $T-$dependence for the Dirac fermionic liquid in a doped graphene sheet~\cite{principi2016bulk}. The analytical relation for $l_{\nu}$ is given by 
\begin{equation}
\label{lnu_theory}
l_{\nu}=\frac{45 v_f (1+N_f \alpha_{ee})^2 \hbar \epsilon_f}{64 \pi \sqrt{2} N_f \alpha_{ee}^2 (k_B T)^2}
 \end{equation}
 where, $N_f=4$ is the number of Fermionic flavours in graphene, $\alpha_{ee}=7.3$ the fine structure constant and $k_B$ the Boltzmann constant. $\alpha_{ee}$ is computed from the relation~\cite{principi2016bulk,kotov2012electron}
 \begin{equation}
\label{alpha_ee}
\alpha_{ee}=\frac{e^2}{\epsilon \hbar v_f}
 \end{equation}
 where $\epsilon$ is the electrical permittivity. For the current work, we have used the permittivity of $h-$BN considering a dielectric constant of 4~\cite{kim2012synthesis}. Eq.~\ref{lnu_theory} (solid line in Fig.~\ref{width_dep}e) has no fitting parameters and matches the extracted values of $l_{\nu}$(closed circles in Fig.~\ref{width_dep}e) for the intermediate temperature range, $178<T<$~300~K (shaded region in Fig.~\ref{width_dep}e). The extracted flow parameters are thus physically relevant only for these intermediate temperatures. Additionally, we note that the assumptions of viscous flow made in the formulation of the transport equations (Eq.~\ref{continuity_eq}~\&~\ref{momentum_conserv_eq}) limit their validity, and as a result, the validity of the geometry dependent analysis for values of $T$ where Poiseuille flow is present. In combination, the physical relevance of the extracted flow parameters and the validity of the viscous flow equations confirm the presence of Poiseuille flow for $178<T<300$~K in $GrFETs$. Finally, this observation is strengthened by the fact that similar $T$ values are obtained for both the current bias measurements (Section~\ref{Current Bias}) and width dependent measurements (Fig.~\ref{fig_gurzhi}e and Fig.~\ref{width_dep}e). We also look at the number density dependence of the flow parameters in Suppementary Fig.~S6. The geometry dependent measurements indicate the presence of hydrodynamic effects for number densities $> 10^{11}$~cm$^{-2}$. Further discussions are provided in Suppementary Fig.~S6.

\begin{figure*}
\includegraphics[scale=0.9]{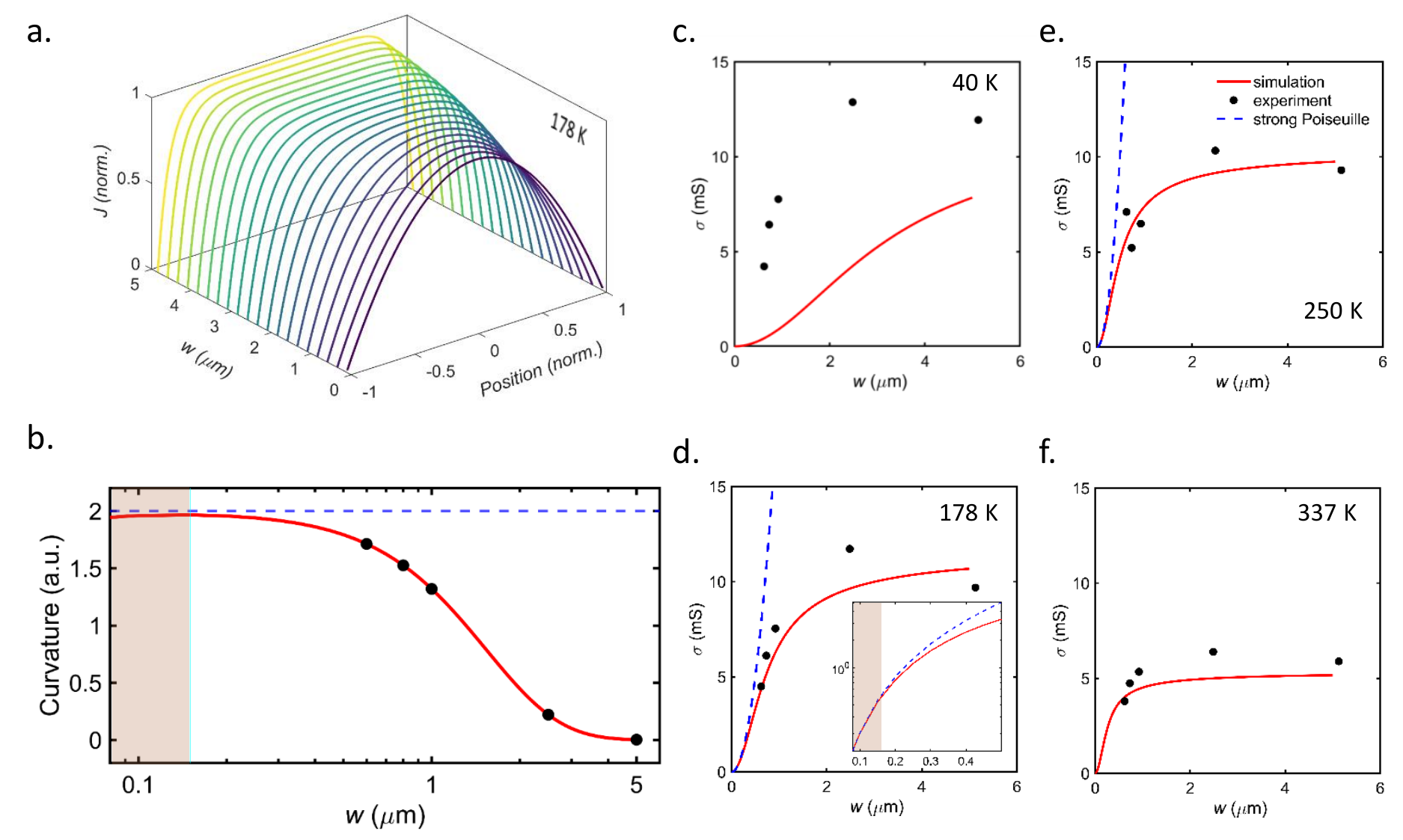}
\caption{\textbf{Finite element modelling of graphene channel.}  a. Plots of $J_{norm}$ as a function of $w$ and $Position~(norm.)$ at $T=178~$K depicting a parabolic current flow profile for small values of $w$ and a uniform $J_{norm}$ for the largest simulated widths. b. Flow curvature (solid line) as a function of $w$ computed from flow profiles in a. at $Position~(norm.)=0$. Filled circles indicate the curvatures for measured values of $w$ in the current work. The dashed line represents curvature for an ideal parabolic flow. Shaded region indicates $w$ values for strong Poiseuille flow (from inset of subsection d.).  c.,d.,e., and f. Experimental (filled circles) and simulated (solid line) values of $\sigma$ as a function of $w$ for $T=40, 178, 250$~and~$337$~K, respectively. Dashed line in subsection d. and e. indicates the $w^2$ scaling of $\sigma$ expected for strong Poiseuille effects. Inset of d. compares  the $w^2$ scaling and simulated $\sigma$ for $w\leq0.5~\mu$m. Shaded region represents the predicted $w$ values for strong Poiseuille flow. The finite element modelling is performed at $n=7\times10^{11}$~cm$^{-2}$.}
\label{simulation}
\end{figure*}
 
 \section{Finite element methods}\label{finite_element}

 Finally, we turn to finite element calculations (implemented in Comsol Multiphysics 5.6) to model the Poiseuille flow in $GrFETs$. This process involves numerically solving Eq.~\ref{continuity_eq} and~\ref{momentum_conserv_eq} with an uniform electric field ($\nabla \phi$) and no-slip boundary conditions for different values of $w$. The input parameters, $D_{\nu}$ and $\sigma_0$, are calculated from the theoretical predictions of $l_\nu$ (Eq.~\ref{lnu_theory}) and the field effect mobility ($\mu$). Our channels are oriented along the x-axis while their width follows the y-axis. An example of a simulated flow profile is depicted in Fig.~\ref{simulation}a for $T=178~$K. We observe strong Poiseuille effects for small channel widths, indicated by the parabolic dependence of the normalized current density (\textbf{$J_{norm}=J_x(y)/J_x(y=0)$}) on the normalized channel position ($Position~(norm.)=y/(w/2)$), and a weak Poiseuille flow, indicated by a constant $J_{norm}$, for larger values of $w$. The curvature of the flow profile computed at the center of the channel ($Position~(norm.)=0$) also exhibits a width dependence (solid line in Fig.~\ref{simulation}b). It saturates to a value of 2 for the smallest widths with a parabolic flow profile ($w=0.1~\mu$m in Fig.~\ref{simulation}a) and reduces to 0 for larger widths with a constant $J_{norm}$ ($w=5~\mu$m in Fig.~\ref{simulation}a). The filled circles in Fig.~\ref{simulation}b correspond to the measured values of $w$ in the current work. Further, we extract the channel conductivity ($\sigma$) from the simulated flow profile. In Fig.~\ref{simulation}c to f, the simulated (solid red line) and experimental (filled circles) values of $\sigma$ are compared for four different values of $T$. The simulated results match experimental data for temperatures in the Poiseuille flow regime ($T=178$ and $250$~K), with mismatches at lower ($T=40$~K) and higher ($T=337$~K) temperatures, where our experiments indicate the absence of hydrodynamic effects.  
 
  In what follows, we quantify our observations and predict the values of $w$ necessary for a strong Poiseuille flow. As discussed before, the strong Poiseuille regime is characterized by a $w^2$ scaling of $\sigma$. This is indicated by the dashed blue lines in Fig.~\ref{simulation}d and e, computed with the same parameter values as in the simulation (solid red line). Indeed, the experimentally measured widths in our $GrFETs$ devices are too large for observing this effect as the flow is in the weak Poiseuille regime. The strong Poiseuille regime is determined by computing the relative difference between the simulated (solid red line) and strong Poiseuille (dashed blue line) values of $\sigma$. Values of $w$ for which the relative difference remains below 10\% are considered to be in the strong Poiseuille regime. The criteria is satisfied for $w<0.15~\mu$m, indicated by the shaded region in the inset of Fig.~\ref{simulation}d. Another indicator of the strong Poiseuille regime is the saturation of the flow curvature to a value of 2 (shaded region in Fig.~\ref{simulation}b). In conclusion, our finite element model confirms our $GrFETs$ are in the weak Poiseuille regime and predicts strong Poiseuille flow for $w<0.15~\mu$m in $GrFETs$. Comparing the measured values of $l_\nu$ (Fig.~\ref{width_dep}e) with the computed length scale for strong hydrodynamics, we would expect this behavior to be present at a slightly higher temperature than the range indicated in the current work in order to satisfy the criteria $l_{ee}<w$.

\section{Discussions and Conclusions}

Strikingly, we observe hydrodynamic effects at different charge number densities in the geometry dependence study ($n>1\times10^{11}$~cm$^{-2}$) (Supplementary Fig.~S6) and the current bias measurements ($1.6\times10^{11} \lesssim n \lesssim 2.4\times10^{11}$~cm$^{-2}$), for which we have no clear, unique explanation. Similar behavior has been seen in previous reports, with current bias measurements showing viscous effects for lower number density ranges compared to negative vicinity resistance~\cite{bandurin2016negative}. This could indicate either a number density selectivity in the transport probes of charge hydrodynamics or a number density variability of the hydrodynamic regime in graphene. 
   
   In conclusion, we demonstrate the electrical signatures of Poiseuille flow in charge transport through graphene channels using the Gurzhi effect and geometry dependent measurements. We observe a negative slope in the differential resistance measurements and a strong geometry dependence of the channel conductivity, both indicative of the presence of viscous effects. The two different measurements types, in conjunction, enable us to define a temperature range for the viscous effects to occur, which reaches as high as 300~K. Moreover, we provide an electrical transport-based characterization framework for detecting Poiseuille flow in conductors that can easily be extended to other conductors. Our experimental observations are corroborated by finite-element calculations that shed light on the channel geometries required for a strong Poiseuille flow. Our findings offer promising prospects for functional hydrodynamic graphene devices in the near future, such as geometric rectifiers like Tesla valves and charge amplifiers based on the electronic Venturi effects~\cite{taubert2011electron,taubert2010electron,szelong2017positive}. 
   
\section{Methods}

\subsection{Device Fabrication}

$GrFETs$ measured in the current work are fabricated using the following process: the heterostructure assembly begins with the mechanical exfoliation of individual layers from bulk crystals (the graphite is obtained commercially from NGS Trading \& Consulting GmbH and the $h$-BN is from our co-authors in National Institute for Materials Science, Japan). Exfoliated flakes are optically screened using an optical microscope (Zeiss Axio Imager M2m). The layer number and quality of grphene flakes are determined using Raman spectroscopy (See Supplementary Fig.~S7). We use a Raman system with a backscattering geometry confocal Raman microscope (WITec, Alpha 300 R). An excitation laser with a wavelength of 532 nm from a diode laser is used for all Raman measurements. The thickness of $h-$BN layers is characterized using atomic force microscopy in the tapping mode (Bruker Icon3 AFM). Selected flakes are aligned and stacked layer by layer using commercially available micromanipulator (hq Graphene 2D Heterostructure Transfer System). The pick-up process is performed using PDMS backed polycarbonate (PC) stamps. PC shows good adherence to $h$-BN for $330~$K$\leq T \leq 350~$K. After successful pick-up of the top $h$-BN layer, the middle graphene layer and bottom $h$-BN layer are picked up using the van der Waals interaction between the different two-dimensional materials. Finally, the heterostructure is deposited on a pre-patterned Si$^{++}$/ SiO$_2$ (285~nm) wafer at $T=450~$K.  The PC is dissolved in dchlorometane (DCM). The transfer process followed is similar to that in Ref.~\citen{zomer2014fast}. Following this, a thermal annealing step is performed at 650~K with H$_2$/Ar 35/200 sccm for 3 hours to improve the quality of the heterostructures by reducing interfacial bubbles, contaminants etc. (See Supplementary Fig.~S8 and S9). The small electrodes are defined using EBL and RIE (CHF$_3 = 40$~sccm, O$_2 = 4$~sccm, $P=60$~mTorr, Power = 60~W). Next, we deposit 5/25 nm Cr/Au directly using ebeam evaporator for edge-contacted electrodes and the pattern is lifted off using acetone for 45 min~\cite{wang2013one}. After that, the second EBL and metals deposition (5/65 nm Cr/Au) are performed for contact pads. Finally, we perform the third ebeam processing, followed by the second RIE for defining the channel shape. The morphology of the device is recorded by the same optical microscope mentioned above. Channel widths of devices are acquired from SEM (Hitachi S-4800). For the width dependent device the channel width ($w$) was varied while keeping the channel length constant.

For the current bias measurements, four devices on four different chips were fabricated. All samples demonstrate a similar scaling of d$V$/d$I$ at base temperature. We observe the current biasing measurements have a tendency to generate irreversible hysteresis and electron doping (shifting of the Dirac point to negative gate voltages) in devices. Due to this we could perform the temperature characterization in one device and partial temperature characterization in a second device.

For the width dependent measurements, the largest four values of $w$ are from a single large graphene channel which was shaped into channels of different widths, while the smallest $w$ value is obtained from a second device on a different chip made following similar fabrication protocols. 
 

\subsection{Measurement Details}
 The four-probe resistance of $GrFETs$ was measured using a lock-in technique, with an AC source-drain current ( of $I_{AC}=100$~nA (231.45~Hz); $I_{DC}=0$) while measuring the resulting voltage drop (Lockins used are SRS 830 or EG\&G 7265). The SRS CS580 voltage controlled current source is used to supply $I_{AC}$ to the sample. The gate voltage is supplied by an AdWin Gold \rom{2} connected to a voltage amplifier (Physics Basel SP 908). From the applied gate voltage, the number density is computed using $n=C_{ox}(V_{g}-V_{d})/e$~\cite{Novoselov_graphene_1}, where, $C_{ox}$ is the oxide capacitance (series combination of SiO$_2$ (285~nm) and bottom $h$-BN), $e$ is the electron charge, $V_g$ and $V_d$ are the gate and the Dirac point voltage, respectively.
  
For the current bias measurements, the DC bias is supplied by an AdWin Gold \rom{2} and the AC bias by a lockin. The two bias voltages are added using a summing amplifier SIM980. The current source CS580 converts the summed voltage to the supplied current bias. An $I_{AC}=100$~nA (231.5~Hz) is used for the measurements and the output voltage from the sample measured using the lockin.

The width dependent measurements follow similar measurement technique as the transfer characteristic measurements. 

The current bias measurements are carried out on a closed cycle manual probe station (Lakeshore). For width dependence measurements, the samples are wire bonded to a chip carrier and measured in a commercially available closed cycle cryostat from Advanced Research Systems (ARS).



\section{Acknowledgement}

 W.H., T.P. and M.C. acknowledge funding from the Swiss National Science Foundation under the Sinergia grant no. 189924 (Hydronics). K.W. and T.T. acknowledge support from the JSPS KAKENHI (Grant Numbers 19H05790 and 20H00354). M.L.P. acknowledges funding from the Swiss National Science Foundation under Spark grant no. 196795. and the Eccellenza Professorial Fellowship no. PCEFP2\textunderscore203663, as well as supported by the Swiss State Secretariat for Education, Research and Innovation (SERI) under contract number MB22.00076. The authors acknowledge support from the Multiphysics Hub @ Empa for the COMSOL Multiphyics calculations. We thank the Cleanroom Operations Team of the Binnig and Rohrer Nanotechnology Center (BRNC) for their help and support. We also thank our project partners Dr.~Bernd Gotsmann, Prof.~Ilaria Zardo, Prof.~Nicola Marzari, Dr.~Ivan Shorubalko for discussions. W.H. would like to thank Prof.~Ilaria Zardo for her mentorship.

\vspace{1cm}

\section{Author contributions statement}

W.H. and T.P. contributed equally to this work. W.H., T.P., M.L.P. and M.C. conceived and designed the experiments. W.H. and T.P. prepared the devices and performed the electrical measurements. W.H. and T.P. performed the annealing, Raman, SEM and AFM measurements.  K.W. and T. T. synthesized the hBN flakes. W.H., T.P., M.L.P., and M.C. did the data analysis. M.L.P. performed the finite element calculations, with the help of T.P.. W.H., T.P., M.L.P., and M.C. discussed the figures and wrote the manuscript. M.L.P. and M.C. supervised the study. All authors discussed the results and implications and commented on the manuscript.

\section{Competing interests}
The authors declare that there are no competing interests.

\bibliographystyle{achemso}
\bibliography{bibliography_9}

\providecommand{\latin}[1]{#1}
\makeatletter
\providecommand{\doi}
  {\begingroup\let\do\@makeother\dospecials
  \catcode`\{=1 \catcode`\}=2 \doi@aux}
\providecommand{\doi@aux}[1]{\endgroup\texttt{#1}}
\makeatother
\providecommand*\mcitethebibliography{\thebibliography}
\csname @ifundefined\endcsname{endmcitethebibliography}
  {\let\endmcitethebibliography\endthebibliography}{}
\begin{mcitethebibliography}{42}
\providecommand*\natexlab[1]{#1}
\providecommand*\mciteSetBstSublistMode[1]{}
\providecommand*\mciteSetBstMaxWidthForm[2]{}
\providecommand*\mciteBstWouldAddEndPuncttrue
  {\def\EndOfBibitem{\unskip.}}
\providecommand*\mciteBstWouldAddEndPunctfalse
  {\let\EndOfBibitem\relax}
\providecommand*\mciteSetBstMidEndSepPunct[3]{}
\providecommand*\mciteSetBstSublistLabelBeginEnd[3]{}
\providecommand*\EndOfBibitem{}
\mciteSetBstSublistMode{f}
\mciteSetBstMaxWidthForm{subitem}{(\alph{mcitesubitemcount})}
\mciteSetBstSublistLabelBeginEnd
  {\mcitemaxwidthsubitemform\space}
  {\relax}
  {\relax}

\bibitem[Ashcroft and Mermin(1976)Ashcroft, and Mermin]{ashcroft1976solid}
Ashcroft,~N.~W.; Mermin,~N.~D. \emph{Solid state physics}; Holt, Rinehart and
  Winston, New York London, 1976\relax
\mciteBstWouldAddEndPuncttrue
\mciteSetBstMidEndSepPunct{\mcitedefaultmidpunct}
{\mcitedefaultendpunct}{\mcitedefaultseppunct}\relax
\EndOfBibitem
\bibitem[Datta(1997)]{datta1997electronic}
Datta,~S. \emph{Electronic transport in mesoscopic systems}; Cambridge
  University Press, 1997\relax
\mciteBstWouldAddEndPuncttrue
\mciteSetBstMidEndSepPunct{\mcitedefaultmidpunct}
{\mcitedefaultendpunct}{\mcitedefaultseppunct}\relax
\EndOfBibitem
\bibitem[Landau and Lifshitz(2013)Landau, and Lifshitz]{landau2013fluid}
Landau,~L.~D.; Lifshitz,~E.~M. \emph{Fluid Mechanics}; Elsevier, 2013;
  Vol.~6\relax
\mciteBstWouldAddEndPuncttrue
\mciteSetBstMidEndSepPunct{\mcitedefaultmidpunct}
{\mcitedefaultendpunct}{\mcitedefaultseppunct}\relax
\EndOfBibitem
\bibitem[Batchelor and Batchelor(2000)Batchelor, and
  Batchelor]{batchelor2000introduction}
Batchelor,~C.~K.; Batchelor,~G. \emph{An introduction to fluid dynamics};
  Cambridge university press, 2000\relax
\mciteBstWouldAddEndPuncttrue
\mciteSetBstMidEndSepPunct{\mcitedefaultmidpunct}
{\mcitedefaultendpunct}{\mcitedefaultseppunct}\relax
\EndOfBibitem
\bibitem[Ho \latin{et~al.}(2018)Ho, Yudhistira, Chakraborty, and
  Adam]{ho2018theoretical}
Ho,~D.~Y.; Yudhistira,~I.; Chakraborty,~N.; Adam,~S. Theoretical determination
  of hydrodynamic window in monolayer and bilayer graphene from scattering
  rates. \emph{Phys. Rev. B} \textbf{2018}, \emph{97}, 121404\relax
\mciteBstWouldAddEndPuncttrue
\mciteSetBstMidEndSepPunct{\mcitedefaultmidpunct}
{\mcitedefaultendpunct}{\mcitedefaultseppunct}\relax
\EndOfBibitem
\bibitem[Bandurin \latin{et~al.}(2016)Bandurin, Torre, Kumar, Shalom, Tomadin,
  Principi, Auton, Khestanova, Novoselov, Grigorieva, Ponomarenko, Geim, and
  Polini]{bandurin2016negative}
Bandurin,~D.; Torre,~I.; Kumar,~R.~K.; Shalom,~M.~B.; Tomadin,~A.;
  Principi,~A.; Auton,~G.; Khestanova,~E.; Novoselov,~K.; Grigorieva,~I.;
  Ponomarenko,~L.~A.; Geim,~A.~K.; Polini,~M. Negative local resistance caused
  by viscous electron backflow in graphene. \emph{Science} \textbf{2016},
  \emph{351}, 1055--1058\relax
\mciteBstWouldAddEndPuncttrue
\mciteSetBstMidEndSepPunct{\mcitedefaultmidpunct}
{\mcitedefaultendpunct}{\mcitedefaultseppunct}\relax
\EndOfBibitem
\bibitem[Moll \latin{et~al.}(2016)Moll, Kushwaha, Nandi, Schmidt, and
  Mackenzie]{moll2016evidence}
Moll,~P.~J.; Kushwaha,~P.; Nandi,~N.; Schmidt,~B.; Mackenzie,~A.~P. Evidence
  for hydrodynamic electron flow in PdCoO2. \emph{Science} \textbf{2016},
  \emph{351}, 1061--1064\relax
\mciteBstWouldAddEndPuncttrue
\mciteSetBstMidEndSepPunct{\mcitedefaultmidpunct}
{\mcitedefaultendpunct}{\mcitedefaultseppunct}\relax
\EndOfBibitem
\bibitem[Gooth \latin{et~al.}(2018)Gooth, Menges, Kumar, S{\"u}$\beta$,
  Shekhar, Sun, Drechsler, Zierold, Felser, and Gotsmann]{gooth2018thermal}
Gooth,~J.; Menges,~F.; Kumar,~N.; S{\"u}$\beta$,~V.; Shekhar,~C.; Sun,~Y.;
  Drechsler,~U.; Zierold,~R.; Felser,~C.; Gotsmann,~B. Thermal and electrical
  signatures of a hydrodynamic electron fluid in tungsten diphosphide.
  \emph{Nat. Commun.} \textbf{2018}, \emph{9}, 1--8\relax
\mciteBstWouldAddEndPuncttrue
\mciteSetBstMidEndSepPunct{\mcitedefaultmidpunct}
{\mcitedefaultendpunct}{\mcitedefaultseppunct}\relax
\EndOfBibitem
\bibitem[De~Jong and Molenkamp(1995)De~Jong, and Molenkamp]{de1995hydrodynamic}
De~Jong,~M.; Molenkamp,~L. Hydrodynamic electron flow in high-mobility wires.
  \emph{Phys. Rev. B} \textbf{1995}, \emph{51}, 13389\relax
\mciteBstWouldAddEndPuncttrue
\mciteSetBstMidEndSepPunct{\mcitedefaultmidpunct}
{\mcitedefaultendpunct}{\mcitedefaultseppunct}\relax
\EndOfBibitem
\bibitem[Keser \latin{et~al.}(2021)Keser, Wang, Klochan, Ho, Tkachenko,
  Tkachenko, Culcer, Adam, Farrer, Ritchie, Sushkov, and
  Hamilton]{PhysRevX.11.031030}
Keser,~A.~C.; Wang,~D.~Q.; Klochan,~O.; Ho,~D. Y.~H.; Tkachenko,~O.~A.;
  Tkachenko,~V.~A.; Culcer,~D.; Adam,~S.; Farrer,~I.; Ritchie,~D.~A.;
  Sushkov,~O.~P.; Hamilton,~A.~R. Geometric Control of Universal Hydrodynamic
  Flow in a Two-Dimensional Electron Fluid. \emph{Phys. Rev. X} \textbf{2021},
  \emph{11}, 031030\relax
\mciteBstWouldAddEndPuncttrue
\mciteSetBstMidEndSepPunct{\mcitedefaultmidpunct}
{\mcitedefaultendpunct}{\mcitedefaultseppunct}\relax
\EndOfBibitem
\bibitem[Oka \latin{et~al.}(2019)Oka, Tajima, Ebisuoka, Hirahara, Watanabe,
  Taniguchi, and Yagi]{PhysRevB.99.035440}
Oka,~T.; Tajima,~S.; Ebisuoka,~R.; Hirahara,~T.; Watanabe,~K.; Taniguchi,~T.;
  Yagi,~R. Ballistic transport experiment detects Fermi surface anisotropy of
  graphene. \emph{Phys. Rev. B} \textbf{2019}, \emph{99}, 035440\relax
\mciteBstWouldAddEndPuncttrue
\mciteSetBstMidEndSepPunct{\mcitedefaultmidpunct}
{\mcitedefaultendpunct}{\mcitedefaultseppunct}\relax
\EndOfBibitem
\bibitem[Rold\'an \latin{et~al.}(2008)Rold\'an, L\'opez-Sancho, and
  Guinea]{PhysRevB.77.115410}
Rold\'an,~R.; L\'opez-Sancho,~M.~P.; Guinea,~F. Effect of electron-electron
  interaction on the Fermi surface topology of doped graphene. \emph{Phys. Rev.
  B} \textbf{2008}, \emph{77}, 115410\relax
\mciteBstWouldAddEndPuncttrue
\mciteSetBstMidEndSepPunct{\mcitedefaultmidpunct}
{\mcitedefaultendpunct}{\mcitedefaultseppunct}\relax
\EndOfBibitem
\bibitem[Johannsen \latin{et~al.}(2013)Johannsen, Ulstrup, Bianchi, Hatch,
  Guan, Mazzola, Hornek{\ae}r, Fromm, Raidel, Seyller, \latin{et~al.}
  others]{johannsen2013electron}
Johannsen,~J.~C.; Ulstrup,~S.; Bianchi,~M.; Hatch,~R.; Guan,~D.; Mazzola,~F.;
  Hornek{\ae}r,~L.; Fromm,~F.; Raidel,~C.; Seyller,~T., \latin{et~al.}
  Electron--phonon coupling in quasi-free-standing graphene. \emph{J. Phys.
  Condens. Matter} \textbf{2013}, \emph{25}, 094001\relax
\mciteBstWouldAddEndPuncttrue
\mciteSetBstMidEndSepPunct{\mcitedefaultmidpunct}
{\mcitedefaultendpunct}{\mcitedefaultseppunct}\relax
\EndOfBibitem
\bibitem[Si \latin{et~al.}(2013)Si, Liu, Duan, and Liu]{si2013first}
Si,~C.; Liu,~Z.; Duan,~W.; Liu,~F. First-principles calculations on the effect
  of doping and biaxial tensile strain on electron-phonon coupling in graphene.
  \emph{Phys. Rev. Lett.} \textbf{2013}, \emph{111}, 196802\relax
\mciteBstWouldAddEndPuncttrue
\mciteSetBstMidEndSepPunct{\mcitedefaultmidpunct}
{\mcitedefaultendpunct}{\mcitedefaultseppunct}\relax
\EndOfBibitem
\bibitem[Calandra and Mauri(2007)Calandra, and Mauri]{calandra2007electron}
Calandra,~M.; Mauri,~F. Electron-phonon coupling and electron self-energy in
  electron-doped graphene: Calculation of angular-resolved photoemission
  spectra. \emph{Phys. Rev. B} \textbf{2007}, \emph{76}, 205411\relax
\mciteBstWouldAddEndPuncttrue
\mciteSetBstMidEndSepPunct{\mcitedefaultmidpunct}
{\mcitedefaultendpunct}{\mcitedefaultseppunct}\relax
\EndOfBibitem
\bibitem[Dean \latin{et~al.}(2010)Dean, Young, Meric, Lee, Wang, Sorgenfrei,
  Watanabe, Taniguchi, Kim, Shepard, and Hone]{Dean_graphene_mobility}
Dean,~C.~R.; Young,~A.~F.; Meric,~I.; Lee,~C.; Wang,~L.; Sorgenfrei,~S.;
  Watanabe,~K.; Taniguchi,~T.; Kim,~P.; Shepard,~K.~L.; Hone,~J. Boron nitride
  substrates for high-quality graphene electronics. \emph{Nat. Nanotechnol.}
  \textbf{2010}, \emph{5}, 722\relax
\mciteBstWouldAddEndPuncttrue
\mciteSetBstMidEndSepPunct{\mcitedefaultmidpunct}
{\mcitedefaultendpunct}{\mcitedefaultseppunct}\relax
\EndOfBibitem
\bibitem[Pellegrino \latin{et~al.}(2016)Pellegrino, Torre, Geim, and
  Polini]{pellegrino2016electron}
Pellegrino,~F.~M.; Torre,~I.; Geim,~A.~K.; Polini,~M. Electron hydrodynamics
  dilemma: Whirlpools or no whirlpools. \emph{Phys. Rev. B} \textbf{2016},
  \emph{94}, 155414\relax
\mciteBstWouldAddEndPuncttrue
\mciteSetBstMidEndSepPunct{\mcitedefaultmidpunct}
{\mcitedefaultendpunct}{\mcitedefaultseppunct}\relax
\EndOfBibitem
\bibitem[Svintsov(2018)]{svintsov2018hydrodynamic}
Svintsov,~D. Hydrodynamic-to-ballistic crossover in Dirac materials.
  \emph{Phys. Rev. B} \textbf{2018}, \emph{97}, 121405\relax
\mciteBstWouldAddEndPuncttrue
\mciteSetBstMidEndSepPunct{\mcitedefaultmidpunct}
{\mcitedefaultendpunct}{\mcitedefaultseppunct}\relax
\EndOfBibitem
\bibitem[Principi \latin{et~al.}(2016)Principi, Vignale, Carrega, and
  Polini]{principi2016bulk}
Principi,~A.; Vignale,~G.; Carrega,~M.; Polini,~M. Bulk and shear viscosities
  of the two-dimensional electron liquid in a doped graphene sheet. \emph{Phys.
  Rev. B} \textbf{2016}, \emph{93}, 125410\relax
\mciteBstWouldAddEndPuncttrue
\mciteSetBstMidEndSepPunct{\mcitedefaultmidpunct}
{\mcitedefaultendpunct}{\mcitedefaultseppunct}\relax
\EndOfBibitem
\bibitem[Torre \latin{et~al.}(2015)Torre, Tomadin, Geim, and
  Polini]{torre2015nonlocal}
Torre,~I.; Tomadin,~A.; Geim,~A.~K.; Polini,~M. Nonlocal transport and the
  hydrodynamic shear viscosity in graphene. \emph{Phys. Rev. B} \textbf{2015},
  \emph{92}, 165433\relax
\mciteBstWouldAddEndPuncttrue
\mciteSetBstMidEndSepPunct{\mcitedefaultmidpunct}
{\mcitedefaultendpunct}{\mcitedefaultseppunct}\relax
\EndOfBibitem
\bibitem[Bandurin \latin{et~al.}(2018)Bandurin, Shytov, Levitov, Kumar,
  Berdyugin, Ben~Shalom, Grigorieva, Geim, and Falkovich]{bandurin2018fluidity}
Bandurin,~D.~A.; Shytov,~A.~V.; Levitov,~L.~S.; Kumar,~R.~K.; Berdyugin,~A.~I.;
  Ben~Shalom,~M.; Grigorieva,~I.~V.; Geim,~A.~K.; Falkovich,~G. Fluidity onset
  in graphene. \emph{Nat. Commun.} \textbf{2018}, \emph{9}, 1--8\relax
\mciteBstWouldAddEndPuncttrue
\mciteSetBstMidEndSepPunct{\mcitedefaultmidpunct}
{\mcitedefaultendpunct}{\mcitedefaultseppunct}\relax
\EndOfBibitem
\bibitem[Ku \latin{et~al.}(2020)Ku, Zhou, Li, Shin, Shi, Burch, Anderson,
  Pierce, Xie, Hamo, \latin{et~al.} others]{ku2020imaging}
Ku,~M.~J.; Zhou,~T.~X.; Li,~Q.; Shin,~Y.~J.; Shi,~J.~K.; Burch,~C.;
  Anderson,~L.~E.; Pierce,~A.~T.; Xie,~Y.; Hamo,~A., \latin{et~al.}  Imaging
  viscous flow of the Dirac fluid in graphene. \emph{Nature} \textbf{2020},
  \emph{583}, 537--541\relax
\mciteBstWouldAddEndPuncttrue
\mciteSetBstMidEndSepPunct{\mcitedefaultmidpunct}
{\mcitedefaultendpunct}{\mcitedefaultseppunct}\relax
\EndOfBibitem
\bibitem[Kumar \latin{et~al.}(2017)Kumar, Bandurin, Pellegrino, Cao, Principi,
  Guo, Auton, Shalom, Ponomarenko, Falkovich, \latin{et~al.}
  others]{kumar2017superballistic}
Kumar,~R.~K.; Bandurin,~D.; Pellegrino,~F.; Cao,~Y.; Principi,~A.; Guo,~H.;
  Auton,~G.; Shalom,~M.~B.; Ponomarenko,~L.~A.; Falkovich,~G., \latin{et~al.}
  Superballistic flow of viscous electron fluid through graphene constrictions.
  \emph{Nat. Phys.} \textbf{2017}, \emph{13}, 1182--1185\relax
\mciteBstWouldAddEndPuncttrue
\mciteSetBstMidEndSepPunct{\mcitedefaultmidpunct}
{\mcitedefaultendpunct}{\mcitedefaultseppunct}\relax
\EndOfBibitem
\bibitem[Sulpizio \latin{et~al.}(2019)Sulpizio, Ella, Rozen, Birkbeck, Perello,
  Dutta, Ben-Shalom, Taniguchi, Watanabe, Holder, \latin{et~al.}
  others]{sulpizio2019visualizing}
Sulpizio,~J.~A.; Ella,~L.; Rozen,~A.; Birkbeck,~J.; Perello,~D.~J.; Dutta,~D.;
  Ben-Shalom,~M.; Taniguchi,~T.; Watanabe,~K.; Holder,~T., \latin{et~al.}
  Visualizing Poiseuille flow of hydrodynamic electrons. \emph{Nature}
  \textbf{2019}, \emph{576}, 75--79\relax
\mciteBstWouldAddEndPuncttrue
\mciteSetBstMidEndSepPunct{\mcitedefaultmidpunct}
{\mcitedefaultendpunct}{\mcitedefaultseppunct}\relax
\EndOfBibitem
\bibitem[Vool \latin{et~al.}(2021)Vool, Hamo, Varnavides, Wang, Zhou, Kumar,
  Dovzhenko, Qiu, Garcia, Pierce, \latin{et~al.} others]{vool2021imaging}
Vool,~U.; Hamo,~A.; Varnavides,~G.; Wang,~Y.; Zhou,~T.~X.; Kumar,~N.;
  Dovzhenko,~Y.; Qiu,~Z.; Garcia,~C.~A.; Pierce,~A.~T., \latin{et~al.}  Imaging
  phonon-mediated hydrodynamic flow in WTe2. \emph{Nat. Phys.} \textbf{2021},
  \emph{17}, 1216--1220\relax
\mciteBstWouldAddEndPuncttrue
\mciteSetBstMidEndSepPunct{\mcitedefaultmidpunct}
{\mcitedefaultendpunct}{\mcitedefaultseppunct}\relax
\EndOfBibitem
\bibitem[Aharon-Steinberg \latin{et~al.}(2022)Aharon-Steinberg, V{\"o}lkl,
  Kaplan, Pariari, Roy, Holder, Wolf, Meltzer, Myasoedov, Huber, Yan,
  Falkovich, Levitov, Hücker, and Zeldov]{aharon2022direct}
Aharon-Steinberg,~A.; V{\"o}lkl,~T.; Kaplan,~A.; Pariari,~A.~K.; Roy,~I.;
  Holder,~T.; Wolf,~Y.; Meltzer,~A.~Y.; Myasoedov,~Y.; Huber,~M.~E.; Yan,~B.;
  Falkovich,~G.; Levitov,~L.~S.; Hücker,~M.; Zeldov,~E. Direct observation of
  vortices in an electron fluid. \emph{Nature} \textbf{2022}, \emph{607},
  74--80\relax
\mciteBstWouldAddEndPuncttrue
\mciteSetBstMidEndSepPunct{\mcitedefaultmidpunct}
{\mcitedefaultendpunct}{\mcitedefaultseppunct}\relax
\EndOfBibitem
\bibitem[Zomer \latin{et~al.}(2014)Zomer, Guimar{\~a}es, Brant, Tombros, and
  Van~Wees]{zomer2014fast}
Zomer,~P.; Guimar{\~a}es,~M.; Brant,~J.; Tombros,~N.; Van~Wees,~B. Fast pick up
  technique for high quality heterostructures of bilayer graphene and hexagonal
  boron nitride. \emph{Appl. Phys. Lett.} \textbf{2014}, \emph{105},
  013101\relax
\mciteBstWouldAddEndPuncttrue
\mciteSetBstMidEndSepPunct{\mcitedefaultmidpunct}
{\mcitedefaultendpunct}{\mcitedefaultseppunct}\relax
\EndOfBibitem
\bibitem[Wang \latin{et~al.}(2013)Wang, Meric, Huang, Gao, Gao, Tran,
  Taniguchi, Watanabe, Campos, Muller, Guo, Kim, Hone, Shepard, and
  Dean]{wang2013one}
Wang,~L.; Meric,~I.; Huang,~P.; Gao,~Q.; Gao,~Y.; Tran,~H.; Taniguchi,~T.;
  Watanabe,~K.; Campos,~L.; Muller,~D.; Guo,~J.; Kim,~P.; Hone,~J.;
  Shepard,~K.~L.; Dean,~C.~R. One-dimensional electrical contact to a
  two-dimensional material. \emph{Science} \textbf{2013}, \emph{342},
  614--617\relax
\mciteBstWouldAddEndPuncttrue
\mciteSetBstMidEndSepPunct{\mcitedefaultmidpunct}
{\mcitedefaultendpunct}{\mcitedefaultseppunct}\relax
\EndOfBibitem
\bibitem[Martin \latin{et~al.}(2008)Martin, Akerman, Ulbricht, Lohmann, Smet,
  Von~Klitzing, and Yacoby]{martin2008observation}
Martin,~J.; Akerman,~N.; Ulbricht,~G.; Lohmann,~T.; Smet,~J.~v.;
  Von~Klitzing,~K.; Yacoby,~A. Observation of electron--hole puddles in
  graphene using a scanning single-electron transistor. \emph{Nat. Phys.}
  \textbf{2008}, \emph{4}, 144--148\relax
\mciteBstWouldAddEndPuncttrue
\mciteSetBstMidEndSepPunct{\mcitedefaultmidpunct}
{\mcitedefaultendpunct}{\mcitedefaultseppunct}\relax
\EndOfBibitem
\bibitem[Xue \latin{et~al.}(2011)Xue, Sanchez-Yamagishi, Bulmash, Jacquod,
  Deshpande, Watanabe, Taniguchi, Jarillo-Herrero, and LeRoy]{xue2011scanning}
Xue,~J.; Sanchez-Yamagishi,~J.; Bulmash,~D.; Jacquod,~P.; Deshpande,~A.;
  Watanabe,~K.; Taniguchi,~T.; Jarillo-Herrero,~P.; LeRoy,~B.~J. Scanning
  tunnelling microscopy and spectroscopy of ultra-flat graphene on hexagonal
  boron nitride. \emph{Nat. Mater.} \textbf{2011}, \emph{10}, 282--285\relax
\mciteBstWouldAddEndPuncttrue
\mciteSetBstMidEndSepPunct{\mcitedefaultmidpunct}
{\mcitedefaultendpunct}{\mcitedefaultseppunct}\relax
\EndOfBibitem
\bibitem[Neto \latin{et~al.}(2009)Neto, Guinea, Peres, Novoselov, and
  Geim]{neto2009electronic}
Neto,~A.~C.; Guinea,~F.; Peres,~N.~M.; Novoselov,~K.~S.; Geim,~A.~K. The
  electronic properties of graphene. \emph{Rev. Mod. Phys.} \textbf{2009},
  \emph{81}, 109\relax
\mciteBstWouldAddEndPuncttrue
\mciteSetBstMidEndSepPunct{\mcitedefaultmidpunct}
{\mcitedefaultendpunct}{\mcitedefaultseppunct}\relax
\EndOfBibitem
\bibitem[Vandecasteele \latin{et~al.}(2010)Vandecasteele, Barreiro, Lazzeri,
  Bachtold, and Mauri]{vandecasteele2010current}
Vandecasteele,~N.; Barreiro,~A.; Lazzeri,~M.; Bachtold,~A.; Mauri,~F.
  Current-voltage characteristics of graphene devices: Interplay between
  Zener-Klein tunneling and defects. \emph{Phys. Rev. B} \textbf{2010},
  \emph{82}, 045416\relax
\mciteBstWouldAddEndPuncttrue
\mciteSetBstMidEndSepPunct{\mcitedefaultmidpunct}
{\mcitedefaultendpunct}{\mcitedefaultseppunct}\relax
\EndOfBibitem
\bibitem[Berdyugin \latin{et~al.}(2022)Berdyugin, Xin, Gao, Slizovskiy, Dong,
  Bhattacharjee, Kumaravadivel, Xu, Ponomarenko, Holwill, \latin{et~al.}
  others]{berdyugin2022out}
Berdyugin,~A.~I.; Xin,~N.; Gao,~H.; Slizovskiy,~S.; Dong,~Z.;
  Bhattacharjee,~S.; Kumaravadivel,~P.; Xu,~S.; Ponomarenko,~L.; Holwill,~M.,
  \latin{et~al.}  Out-of-equilibrium criticalities in graphene superlattices.
  \emph{Science} \textbf{2022}, \emph{375}, 430--433\relax
\mciteBstWouldAddEndPuncttrue
\mciteSetBstMidEndSepPunct{\mcitedefaultmidpunct}
{\mcitedefaultendpunct}{\mcitedefaultseppunct}\relax
\EndOfBibitem
\bibitem[Gurzhi(1963)]{gurzhi1963minimum}
Gurzhi,~R. Minimum of resistance in impurity-free conductors. \emph{J Exp Theor
  Phys} \textbf{1963}, \emph{17}, 521\relax
\mciteBstWouldAddEndPuncttrue
\mciteSetBstMidEndSepPunct{\mcitedefaultmidpunct}
{\mcitedefaultendpunct}{\mcitedefaultseppunct}\relax
\EndOfBibitem
\bibitem[Giuliani and Vignale(2005)Giuliani, and Vignale]{giuliani2005quantum}
Giuliani,~G.; Vignale,~G. \emph{Quantum theory of the electron liquid};
  Cambridge University Press, 2005\relax
\mciteBstWouldAddEndPuncttrue
\mciteSetBstMidEndSepPunct{\mcitedefaultmidpunct}
{\mcitedefaultendpunct}{\mcitedefaultseppunct}\relax
\EndOfBibitem
\bibitem[Razeghi(2006)]{razeghi2006fundamentals}
Razeghi,~M. \emph{Fundamentals of solid state engineering}; Springer,
  2006\relax
\mciteBstWouldAddEndPuncttrue
\mciteSetBstMidEndSepPunct{\mcitedefaultmidpunct}
{\mcitedefaultendpunct}{\mcitedefaultseppunct}\relax
\EndOfBibitem
\bibitem[Kotov \latin{et~al.}(2012)Kotov, Uchoa, Pereira, Guinea, and
  Neto]{kotov2012electron}
Kotov,~V.~N.; Uchoa,~B.; Pereira,~V.~M.; Guinea,~F.; Neto,~A.~C.
  Electron-electron interactions in graphene: Current status and perspectives.
  \emph{Rev. Mod. Phys.} \textbf{2012}, \emph{84}, 1067\relax
\mciteBstWouldAddEndPuncttrue
\mciteSetBstMidEndSepPunct{\mcitedefaultmidpunct}
{\mcitedefaultendpunct}{\mcitedefaultseppunct}\relax
\EndOfBibitem
\bibitem[Kim \latin{et~al.}(2012)Kim, Hsu, Jia, Kim, Shi, Dresselhaus,
  Palacios, and Kong]{kim2012synthesis}
Kim,~K.~K.; Hsu,~A.; Jia,~X.; Kim,~S.~M.; Shi,~Y.; Dresselhaus,~M.;
  Palacios,~T.; Kong,~J. Synthesis and characterization of hexagonal boron
  nitride film as a dielectric layer for graphene devices. \emph{ACS nano}
  \textbf{2012}, \emph{6}, 8583--8590\relax
\mciteBstWouldAddEndPuncttrue
\mciteSetBstMidEndSepPunct{\mcitedefaultmidpunct}
{\mcitedefaultendpunct}{\mcitedefaultseppunct}\relax
\EndOfBibitem
\bibitem[Taubert \latin{et~al.}(2011)Taubert, Schinner, Tomaras, Tranitz,
  Wegscheider, and Ludwig]{taubert2011electron}
Taubert,~D.; Schinner,~G.~J.; Tomaras,~C.; Tranitz,~H.-P.; Wegscheider,~W.;
  Ludwig,~S. An electron jet pump: The Venturi effect of a Fermi liquid.
  \emph{J. Appl. Phys.} \textbf{2011}, \emph{109}, 102412\relax
\mciteBstWouldAddEndPuncttrue
\mciteSetBstMidEndSepPunct{\mcitedefaultmidpunct}
{\mcitedefaultendpunct}{\mcitedefaultseppunct}\relax
\EndOfBibitem
\bibitem[Taubert \latin{et~al.}(2010)Taubert, Schinner, Tranitz, Wegscheider,
  Tomaras, Kehrein, and Ludwig]{taubert2010electron}
Taubert,~D.; Schinner,~G.; Tranitz,~H.; Wegscheider,~W.; Tomaras,~C.;
  Kehrein,~S.; Ludwig,~S. Electron-avalanche amplifier based on the electronic
  Venturi effect. \emph{Phys. Rev. B} \textbf{2010}, \emph{82}, 161416\relax
\mciteBstWouldAddEndPuncttrue
\mciteSetBstMidEndSepPunct{\mcitedefaultmidpunct}
{\mcitedefaultendpunct}{\mcitedefaultseppunct}\relax
\EndOfBibitem
\bibitem[Szelong \latin{et~al.}(2017)Szelong, Ludwig, Wieck, and
  Kunze]{szelong2017positive}
Szelong,~M.; Ludwig,~A.; Wieck,~A.~D.; Kunze,~U. Positive centre voltage in
  T-branch junctions on n-type GaAs/AlGaAs based on hydrodynamics.
  \emph{Semicond. Sci. Technol.} \textbf{2017}, \emph{32}, 105005\relax
\mciteBstWouldAddEndPuncttrue
\mciteSetBstMidEndSepPunct{\mcitedefaultmidpunct}
{\mcitedefaultendpunct}{\mcitedefaultseppunct}\relax
\EndOfBibitem
\bibitem[Novoselov \latin{et~al.}(2004)Novoselov, Geim, Morozov, Jiang, Zhang,
  Dubonos, Grigorieva, and Firsov]{Novoselov_graphene_1}
Novoselov,~K.~S.; Geim,~A.~K.; Morozov,~S.~V.; Jiang,~D.; Zhang,~Y.;
  Dubonos,~S.~V.; Grigorieva,~I.~V.; Firsov,~A.~A. Electric Field Effect in
  Atomically Thin Carbon Films. \emph{Science} \textbf{2004}, \emph{306},
  666--669\relax
\mciteBstWouldAddEndPuncttrue
\mciteSetBstMidEndSepPunct{\mcitedefaultmidpunct}
{\mcitedefaultendpunct}{\mcitedefaultseppunct}\relax
\EndOfBibitem
\end{mcitethebibliography}


\providecommand{\latin}[1]{#1}
\makeatletter
\providecommand{\doi}
  {\begingroup\let\do\@makeother\dospecials
  \catcode`\{=1 \catcode`\}=2 \doi@aux}
\providecommand{\doi@aux}[1]{\endgroup\texttt{#1}}
\makeatother
\providecommand*\mcitethebibliography{\thebibliography}
\csname @ifundefined\endcsname{endmcitethebibliography}
  {\let\endmcitethebibliography\endthebibliography}{}
\begin{mcitethebibliography}{5}
\providecommand*\natexlab[1]{#1}
\providecommand*\mciteSetBstSublistMode[1]{}
\providecommand*\mciteSetBstMaxWidthForm[2]{}
\providecommand*\mciteBstWouldAddEndPuncttrue
  {\def\EndOfBibitem{\unskip.}}
\providecommand*\mciteBstWouldAddEndPunctfalse
  {\let\EndOfBibitem\relax}
\providecommand*\mciteSetBstMidEndSepPunct[3]{}
\providecommand*\mciteSetBstSublistLabelBeginEnd[3]{}
\providecommand*\EndOfBibitem{}
\mciteSetBstSublistMode{f}
\mciteSetBstMaxWidthForm{subitem}{(\alph{mcitesubitemcount})}
\mciteSetBstSublistLabelBeginEnd
  {\mcitemaxwidthsubitemform\space}
  {\relax}
  {\relax}

\bibitem[Terr{\'e}s \latin{et~al.}(2016)Terr{\'e}s, Chizhova, Libisch, Peiro,
  J{\"o}rger, Engels, Girschik, Watanabe, Taniguchi, Rotkin, \latin{et~al.}
  others]{terres2016size}
Terr{\'e}s,~B.; Chizhova,~L.; Libisch,~F.; Peiro,~J.; J{\"o}rger,~D.;
  Engels,~S.; Girschik,~A.; Watanabe,~K.; Taniguchi,~T.; Rotkin,~S.,
  \latin{et~al.}  Size quantization of Dirac fermions in graphene
  constrictions. \emph{Nat. Commun.} \textbf{2016}, \emph{7}, 1--7\relax
\mciteBstWouldAddEndPuncttrue
\mciteSetBstMidEndSepPunct{\mcitedefaultmidpunct}
{\mcitedefaultendpunct}{\mcitedefaultseppunct}\relax
\EndOfBibitem
\bibitem[van Wees \latin{et~al.}(1988)van Wees, van Houten, Beenakker,
  Williamson, Kouwenhoven, van~der Marel, and Foxon]{wees1988quantized}
van Wees,~B.; van Houten,~H.; Beenakker,~C. W.~J.; Williamson,~J.~G.;
  Kouwenhoven,~L.~P.; van~der Marel,~D.; Foxon,~C.~T. Quantized Conductance of
  Point Contacts in a Two-Dimensional Electron Gas. \emph{Phys. Rev. Lett.}
  \textbf{1988}, \emph{60}, 848\relax
\mciteBstWouldAddEndPuncttrue
\mciteSetBstMidEndSepPunct{\mcitedefaultmidpunct}
{\mcitedefaultendpunct}{\mcitedefaultseppunct}\relax
\EndOfBibitem
\bibitem[Bandurin \latin{et~al.}(2016)Bandurin, Torre, Kumar, Shalom, Tomadin,
  Principi, Auton, Khestanova, Novoselov, Grigorieva, Ponomarenko, Geim, and
  Polini]{bandurin2016negative}
Bandurin,~D.; Torre,~I.; Kumar,~R.~K.; Shalom,~M.~B.; Tomadin,~A.;
  Principi,~A.; Auton,~G.; Khestanova,~E.; Novoselov,~K.; Grigorieva,~I.;
  Ponomarenko,~L.~A.; Geim,~A.~K.; Polini,~M. Negative local resistance caused
  by viscous electron backflow in graphene. \emph{Science} \textbf{2016},
  \emph{351}, 1055--1058\relax
\mciteBstWouldAddEndPuncttrue
\mciteSetBstMidEndSepPunct{\mcitedefaultmidpunct}
{\mcitedefaultendpunct}{\mcitedefaultseppunct}\relax
\EndOfBibitem
\bibitem[De~Jong and Molenkamp(1995)De~Jong, and Molenkamp]{de1995hydrodynamic}
De~Jong,~M.; Molenkamp,~L. Hydrodynamic electron flow in high-mobility wires.
  \emph{Phys. Rev. B} \textbf{1995}, \emph{51}, 13389\relax
\mciteBstWouldAddEndPuncttrue
\mciteSetBstMidEndSepPunct{\mcitedefaultmidpunct}
{\mcitedefaultendpunct}{\mcitedefaultseppunct}\relax
\EndOfBibitem
\bibitem[Ferrari \latin{et~al.}(2006)Ferrari, Meyer, Scardaci, Casiraghi,
  Lazzeri, Mauri, Piscanec, Jiang, Novoselov, Roth, \latin{et~al.}
  others]{ferrari2006raman}
Ferrari,~A.~C.; Meyer,~J.~C.; Scardaci,~V.; Casiraghi,~C.; Lazzeri,~M.;
  Mauri,~F.; Piscanec,~S.; Jiang,~D.; Novoselov,~K.~S.; Roth,~S.,
  \latin{et~al.}  Raman spectrum of graphene and graphene layers. \emph{Phys.
  Rev. Lett.} \textbf{2006}, \emph{97}, 187401\relax
\mciteBstWouldAddEndPuncttrue
\mciteSetBstMidEndSepPunct{\mcitedefaultmidpunct}
{\mcitedefaultendpunct}{\mcitedefaultseppunct}\relax
\EndOfBibitem
\end{mcitethebibliography}

\end{document}





\title{Supplementary materials for\\
Electronic Poiseuille Flow in Hexagonal Boron Nitride Encapsulated Graphene FETs}

\author{Wenhao Huang}
\thanks{Equal contribution}
\email{}
\affiliation{Empa, Swiss Federal Laboratories for Materials Science and Technology, Transport at Nanoscale Interfaces Laboratory, Überlandstrasse 129, CH-8600 Dübendorf, Switzerland.}
\affiliation{Department of Physics, University of Basel, Klingelbergstrasse 82, CH-4056 Basel, Switzerland.}

\author{Tathagata Paul}
\thanks{Equal contribution}
\email{tathagata.paul@empa.ch}
\affiliation{Empa, Swiss Federal Laboratories for Materials Science and Technology, Transport at Nanoscale Interfaces Laboratory, Überlandstrasse 129, CH-8600 Dübendorf, Switzerland.}

\author{Kenji Watanabe}
\affiliation{Research Center for Functional Materials, National Institute for Materials Science, 1-1 Namiki, Tsukuba, Ibaraki 305-0044, Japan.}

\author{Takashi Taniguchi}
\affiliation{International Center for Materials Nanoarchitectonics, National Institute for Materials Science, 1-1 Namiki, Tsukuba, Ibaraki 305-0044, Japan.}


\author{Mickael L. Perrin}
\email{Mickael.Perrin@empa.ch}
\affiliation{Empa, Swiss Federal Laboratories for Materials Science and Technology, Transport at Nanoscale Interfaces Laboratory, Überlandstrasse 129, CH-8600 Dübendorf, Switzerland.}
\affiliation{Department of Information Technology and Electrical Engineering, ETH Zürich, 8092 Zürich, Switzerland.}

\author{Michel Calame}
\email{michel.calame@empa.ch}
\affiliation{Empa, Swiss Federal Laboratories for Materials Science and Technology, Transport at Nanoscale Interfaces Laboratory, Überlandstrasse 129, CH-8600 Dübendorf, Switzerland.}
\affiliation{Department of Physics, University of Basel, Klingelbergstrasse 82, CH-4056 Basel, Switzerland.}
\affiliation{Swiss Nanoscience Institute, University of Basel, Klingelbergstrasse 82, CH-4056 Basel, Switzerland.}

\maketitle

\newpage
\setcounter{figure}{0}
\renewcommand{\figurename}{Fig.}
\renewcommand{\thefigure}{S\arabic{figure}}

\begin{figure*}
\includegraphics[scale=0.6]{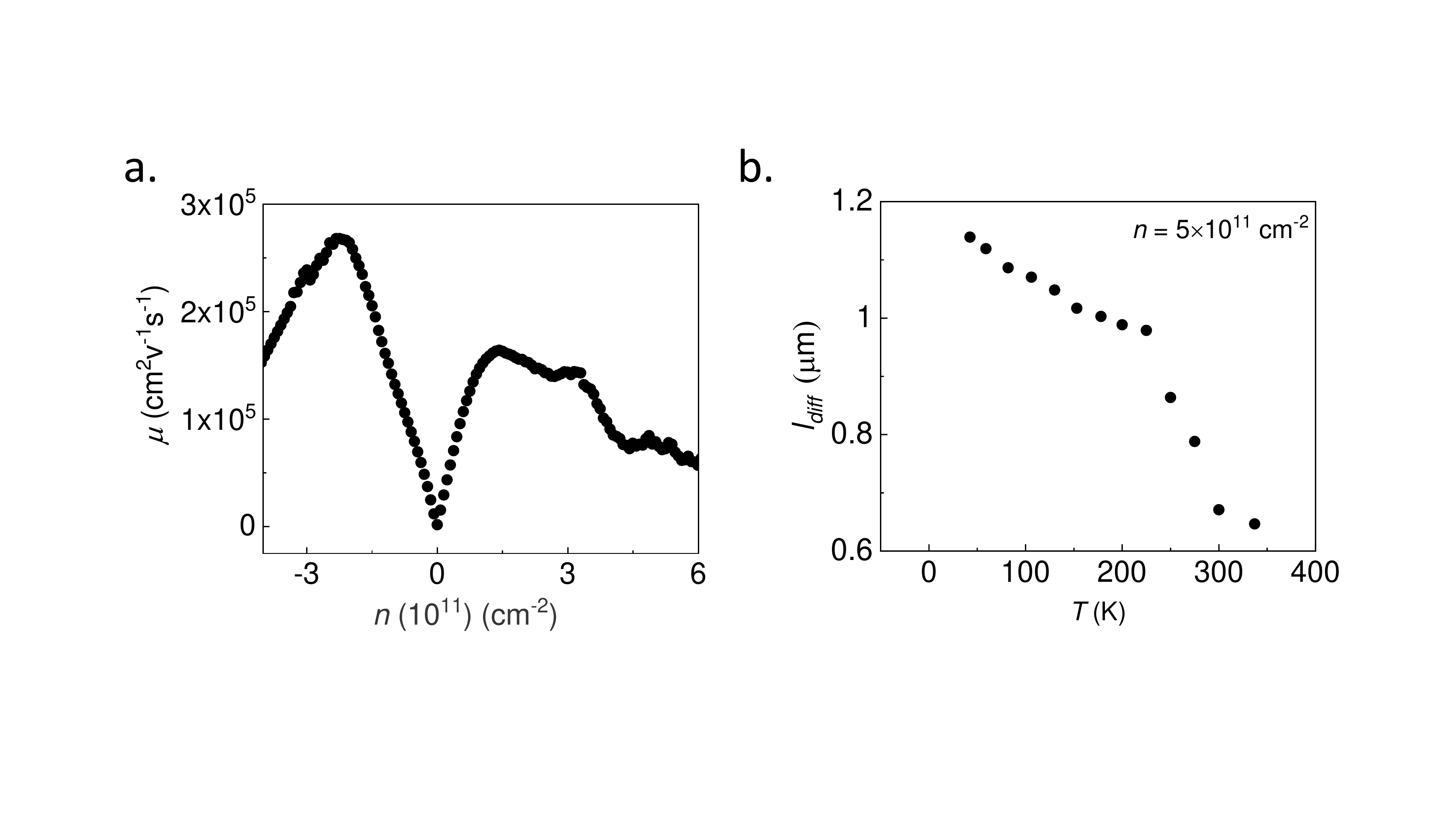}
\caption{\textbf{Mobility and $l_{diff}$ of $GrFETs$} a. Mobility ($\mu$) of $GrFET$ extracted from the transfer characteristic in Fig.~1b of the main text at $T=40$~K. The mobility is extracted from the channel conductance following the Drude relationship  $\sigma = n e \mu$, where $n$ is the number density and $e$ the electronic charge. b. Temperature dependence of $l_{diff}$ extracted from the transfer characteristics following Eq.~1 of the main text. Values of $l_{diff}$ exceeding 1~$\mu$m indicate the quality of our samples.}

\label{Figure.S1}
\end{figure*}

\begin{figure*}
\includegraphics[scale=0.65]{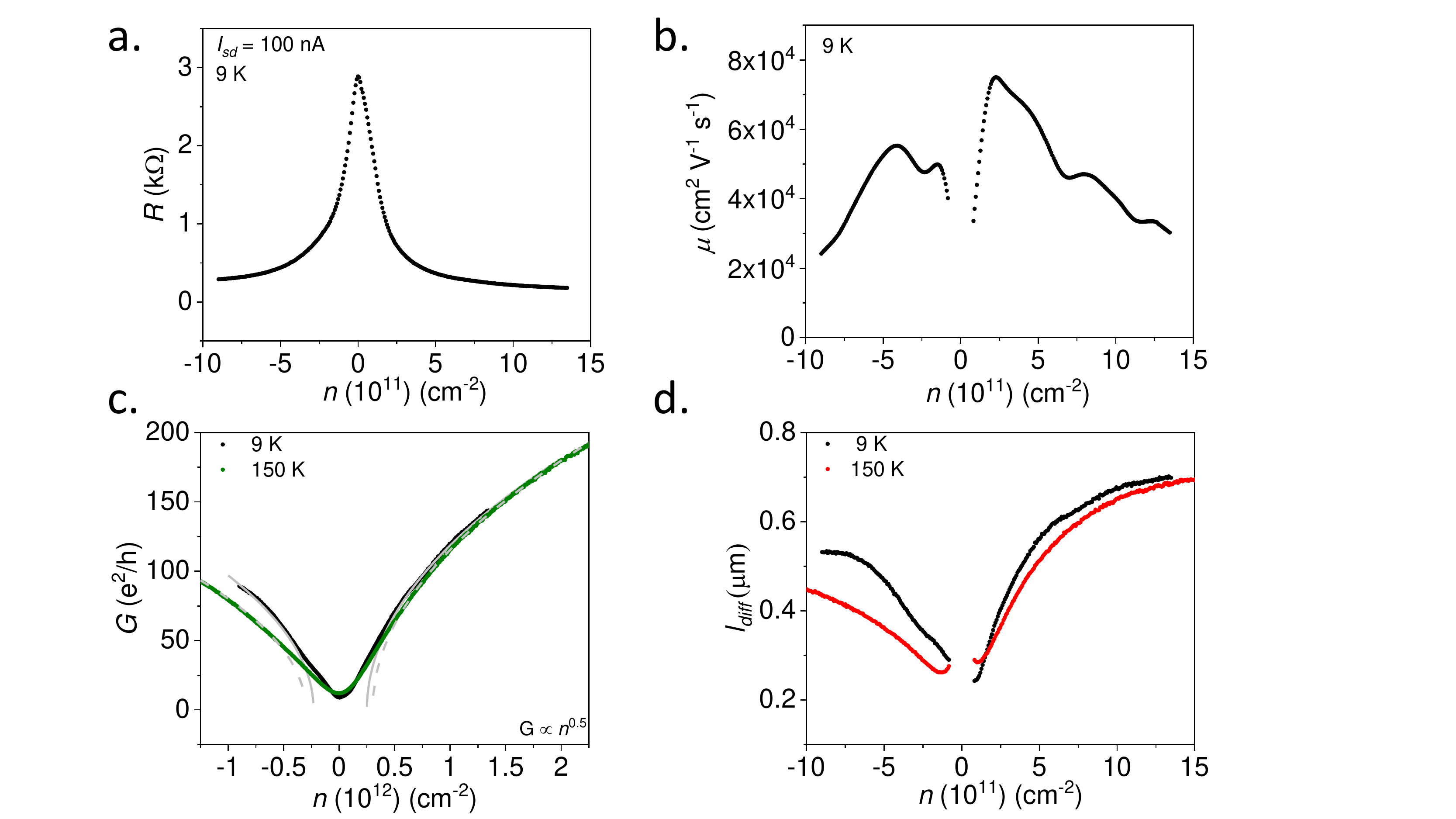}
\caption{\textbf{Basic performance of the measured Gurzhi device.} \textbf{a)} Transfer characteristics of the device used for current bias measurements. Resistance was measured using similar technique as mentioned in the Methods section. \textbf{b)} The mobility ($\mu$) extracted from the transfer characteristics in subsection a. at 9~K. We find a peak mobility of 46000 cm$^{2}$V$^{-1}$s$^{-1}$ at $n \sim 1.9 \times 10^{11} cm^{-2}$ in this device. c. Channel conductance ($G$) as a function of number density at two different temperatures. We observe a $\sqrt n$ scaling of $G$ (gray solid and dashed line for $T=$ 9 and 150~K, respectively), indicative of  the presence of quasi-ballistic transport in the graphene channel~\cite{terres2016size,wees1988quantized}. Additionally, in subsection d., the extracted values of $l_{diff} \gtrsim w$ ($w \sim$0.55~$\mu$m) for temperatures upto 150~K hints towards the presence of quasi-ballistic charge flow. The $l_{diff}$ values are extracted from the transfer characteristics following Eq.~1 of the main text. }

\label{Figure.S2}
\end{figure*}

\begin{figure*}
\includegraphics[scale=0.5]{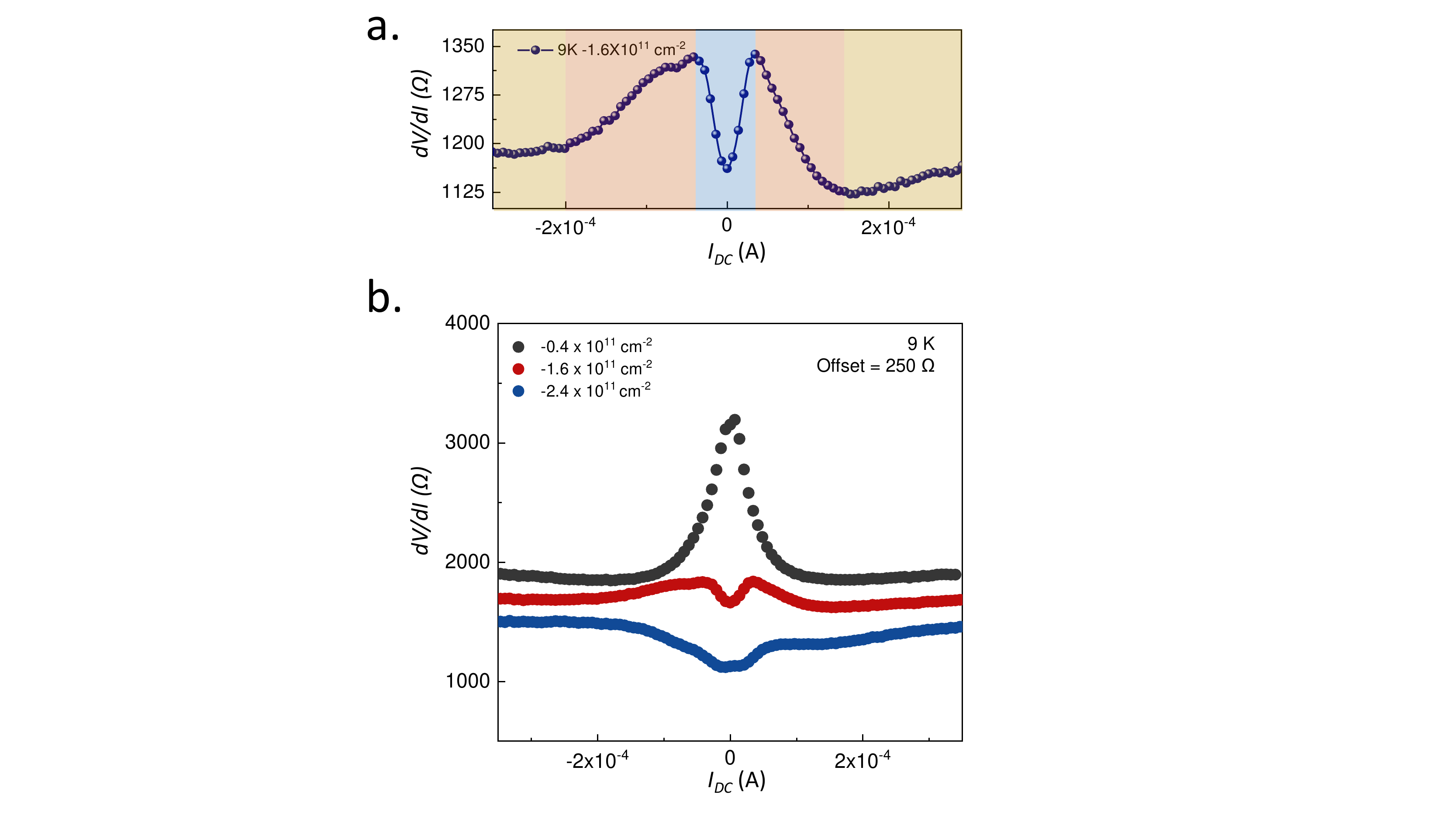}
\caption{\textbf{Differential resistance measurements for hole transport.} \textbf{a)} Plot of d$V$/d$I$ vs $I_{DC}$ for $n =-1.6\times 10^{11}$cm$^{-2}$ at 9 K. The yellow, pink, and blue colored regions indicate the diffusive, Poiseuille, and Knudsen scaling of differential resistance, respectively. \textbf{b)} $I_{DC}$ dependence of differential resistance showing three distinct behaviors similar to those presented in Fig.~2d of the main text~\cite{bandurin2016negative,de1995hydrodynamic}.}

\label{Figure.S3}
\end{figure*}

\begin{figure*}
\includegraphics[scale=0.5]{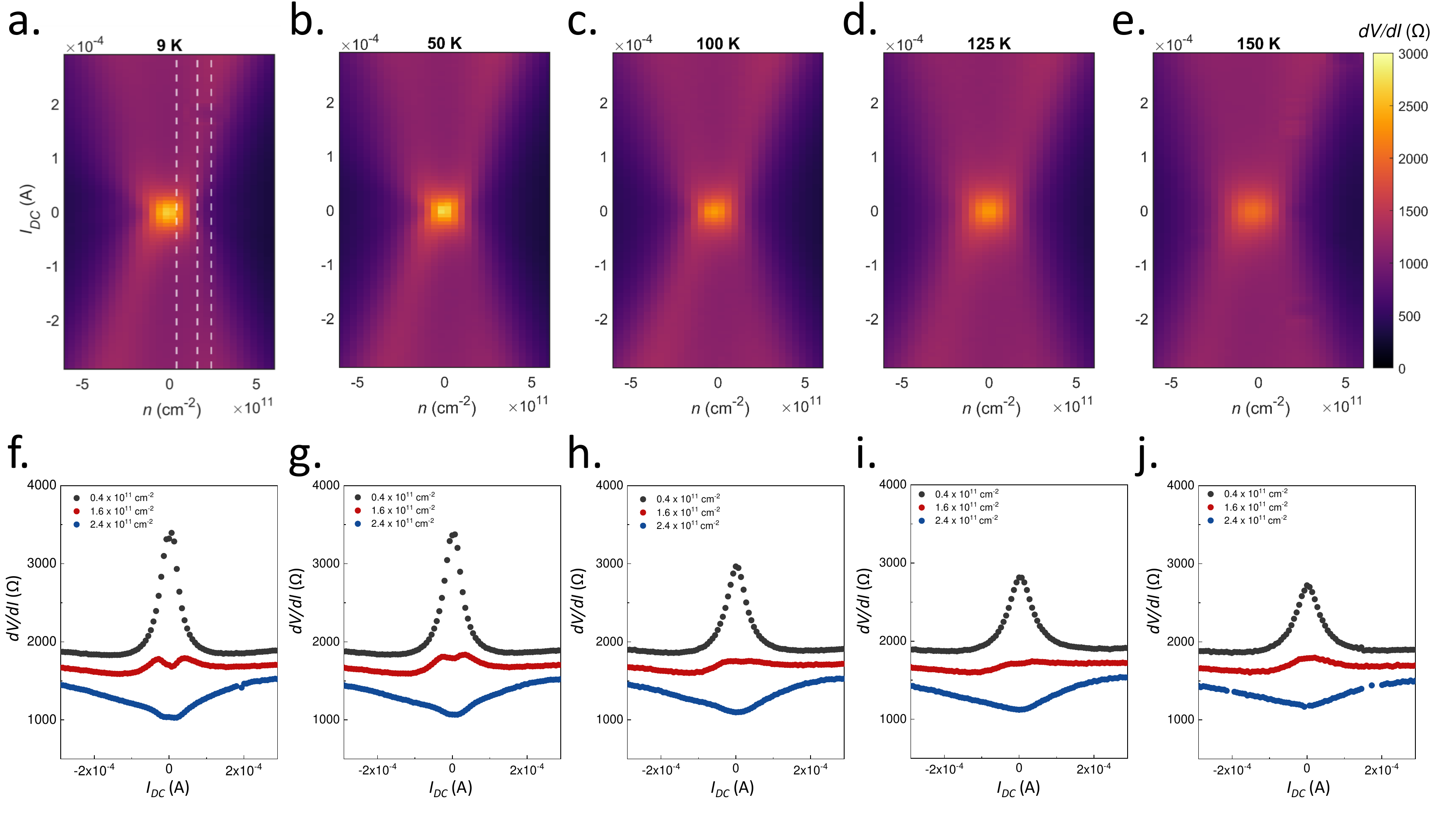}
\caption{\textbf{Temperature dependence of differential resistance measurements.} \textbf{a)}- \textbf{e)} Color plots of dV /dI as a function of $I_{DC}$ and number density ($n$) at different temperatures (from 9~K to 150~K). \textbf{f)}- \textbf{j)} Plots depicting the $I_{DC}$ dependence of differential resistance extracted for the same $n$ values as in Fig.~2d of the main text (white dashed lines for $T=9$~K data) at different temperatures (from 9~K to 150~K)~\cite{bandurin2016negative,de1995hydrodynamic}.}
\label{Figure.S4}
\end{figure*}

\begin{figure*}
\includegraphics[scale=0.65]{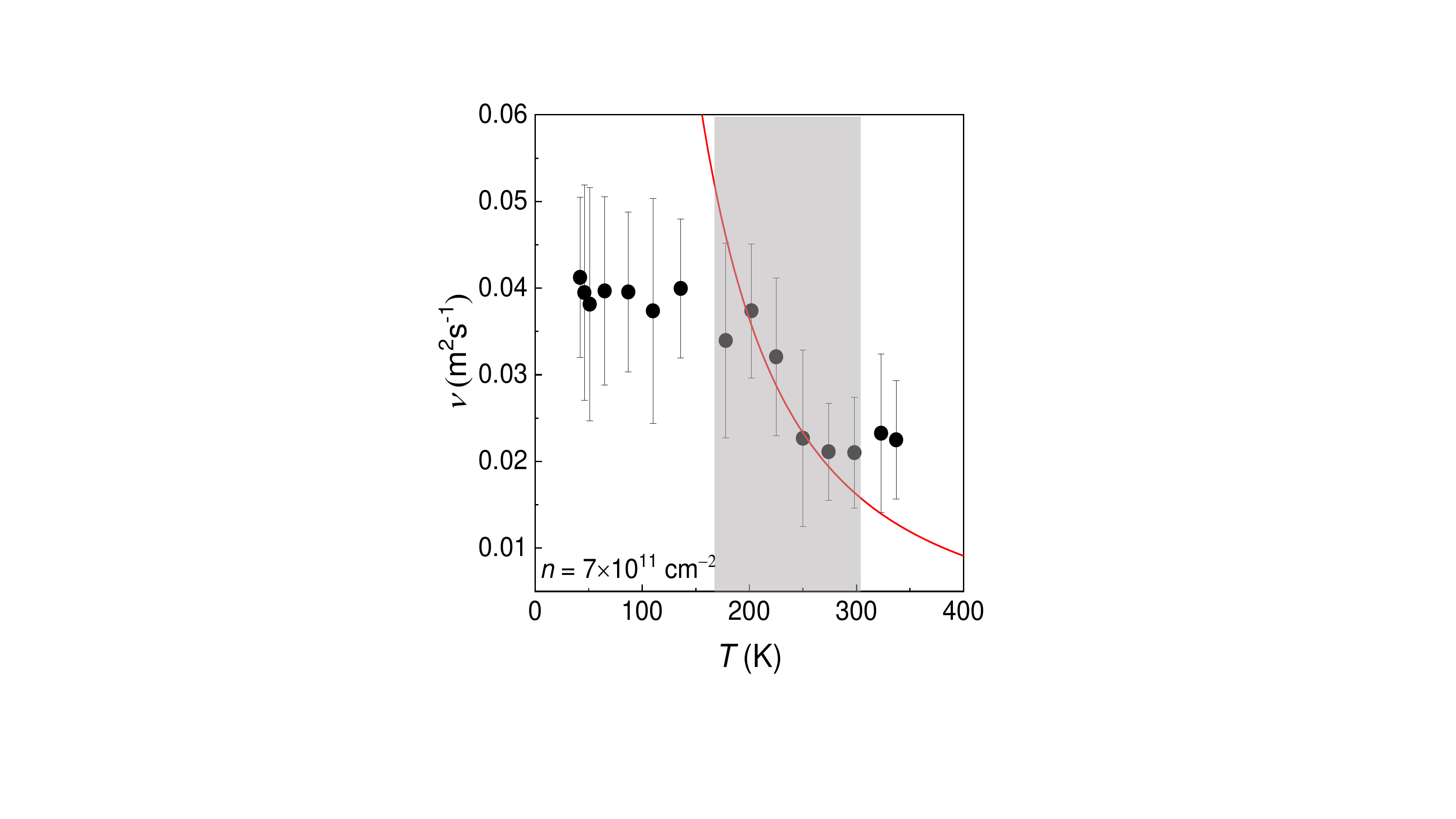}
\caption{\textbf{Temperature dependence of  Viscosity $\nu$.} Viscosity $\nu$ as a function of temperature extracted from the fits (Fig.~3b of main text). Solid line and shaded region represent theoretically predicted scaling of $\nu$, following Eq.7 in main text and the temperature range for Poiseuille efffects in $GrFETs$ from geometry dependent measurements, respectively.}
\label{Figure.S5}
\end{figure*}

\begin{figure*}
\includegraphics[scale=0.55]{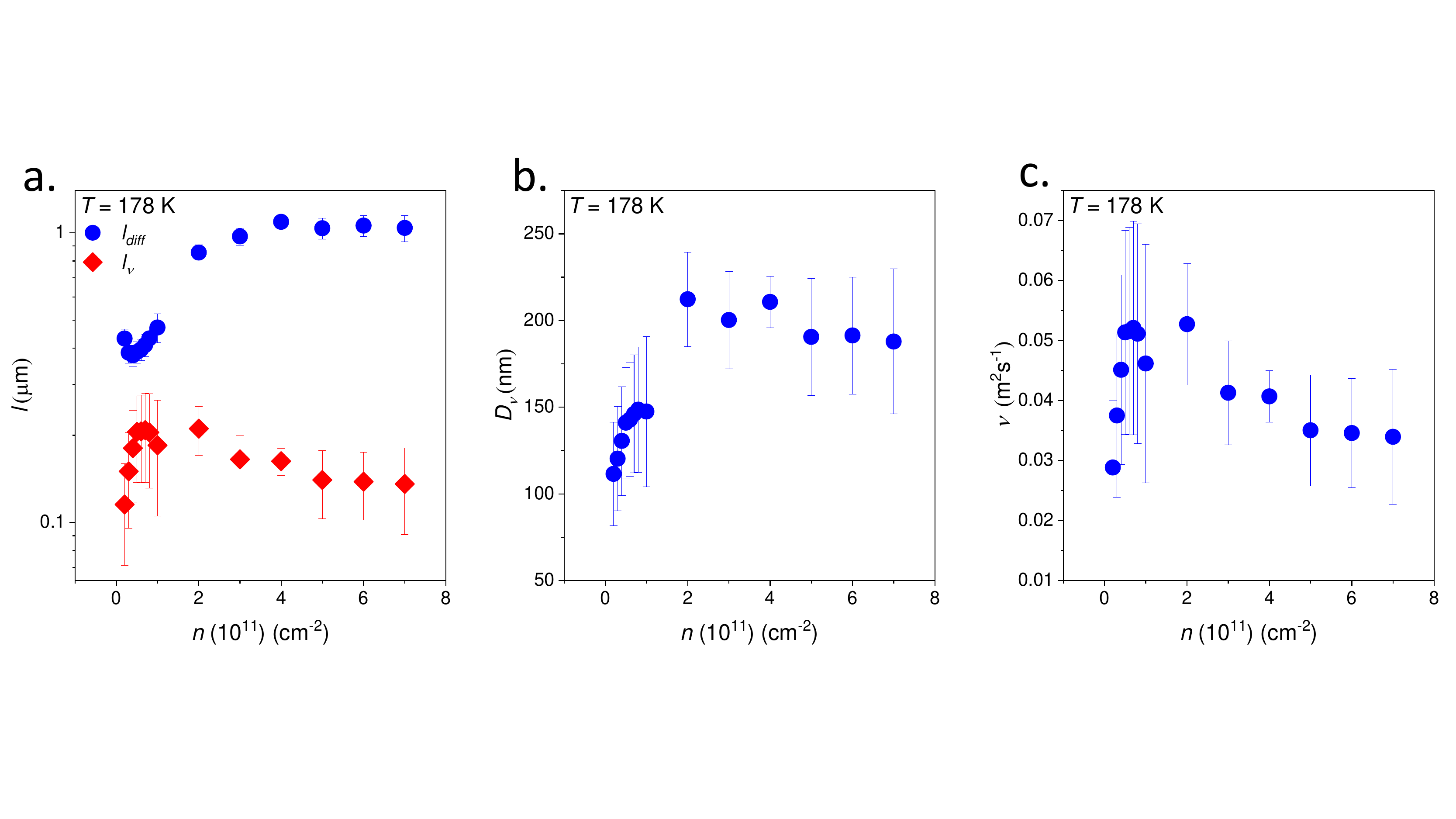}
\caption{\textbf{Number density ($n$) dependence of flow parameters.} \textbf{a)} Plots of diffusive $l_{diff}$ and viscous $l_{\nu}$ mean free path as a function of number density. The ratio $l_{diff} / l_{\nu}$ 
has a value $\sim 10$ at $n=7 \times 10^{11}~$cm$^{-2}$ and reduces to $\sim 2$ at $n=1 \times 10^{11}~$cm$^{-2}$, which indicates the lower bound of number density for charge hydrodynamic effects in $GrFETs$. Number densities higher than $7 \times 10^{11}~$cm$^{-2}$ could not be explored in the current work due to gating limitations and hence, we cannot comment on the upper number density bound of the hydrodynamic regime. \textbf{b)} and \textbf{c)} demonstrate the $n$ dependence of the Gurzhi length ($D_{\nu}$) and kinematic viscosity ($\nu$), respectively. Both parameters demonstrate an enhancement with decreasing number density (for $n>1 \times 10^{11}~$cm$^{-2}$) which is similar to trends observed in previous measurements~\cite{bandurin2016negative}. Values for $n<1 \times 10^{11}~$cm$^{-2}$ have a large error since the sample is no longer in the hydrodynamic regime.}
\label{Figure.S6}
\end{figure*}

\begin{figure*}
\includegraphics[scale=0.55]{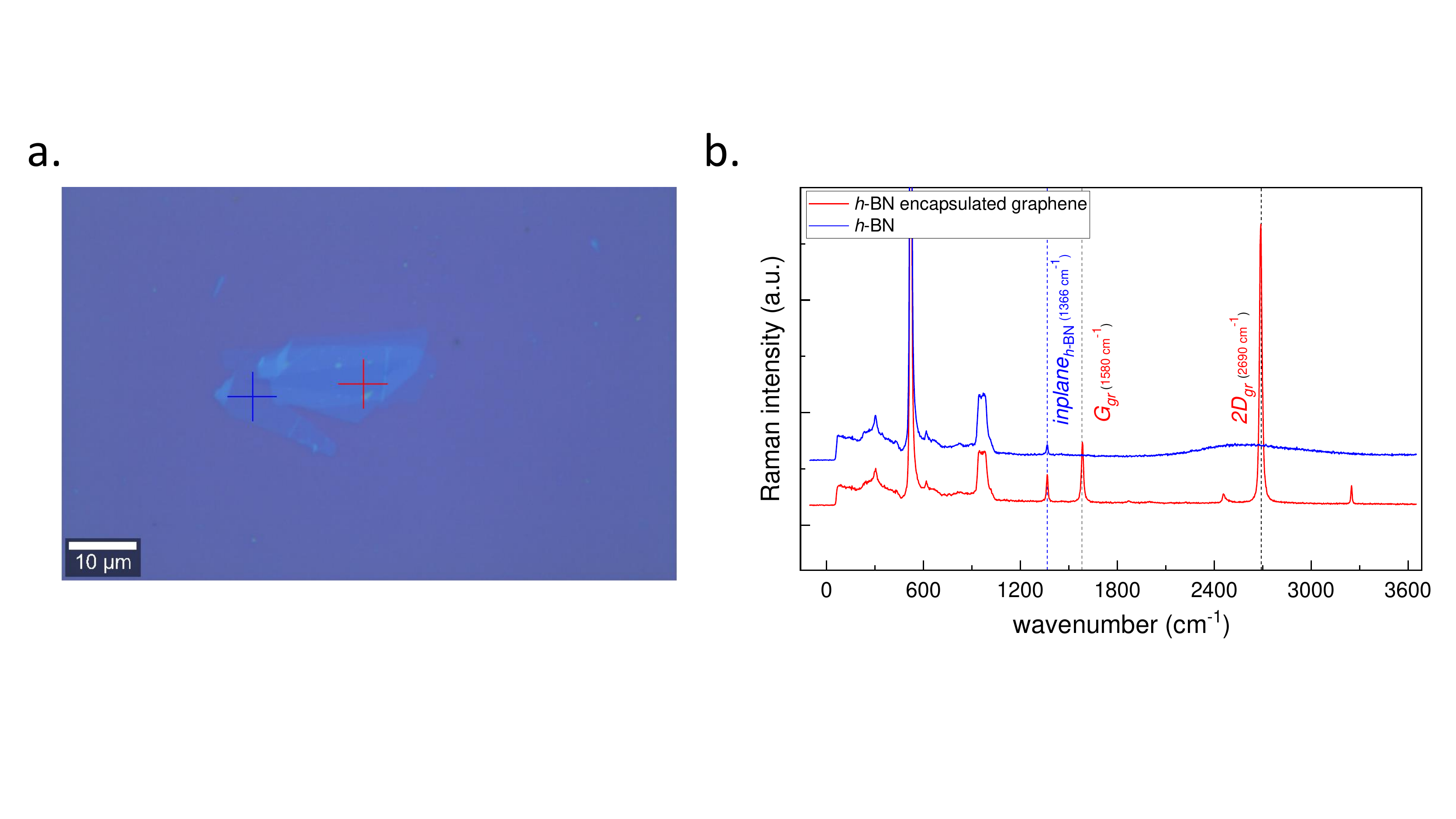}
\caption{\textbf{Raman measurements on $h$-BN encapsulated graphene heterostructure.} \textbf{a)} Optical image of a $GrFET$ device used for Raman characterization. \textbf{b)} Raman spectrums recorded on $h$-BN encapsulated graphene region (red mark in \textbf{a)}) and bare $h$-BN region (blue mark in \textbf{a)}), respectively.  Characteristic G (1580~cm$^{-1}$) and 2D (2690~cm$^{-1}$) peaks of graphene are visible in the encapsulated region with an absence of the defect peak (1350~cm$^{-1}$) ascertaining the crystalline quality of graphene. Furthermore, a higher intensity of the 2D peak when compared to the G peak indicates single layer graphene~\cite{ferrari2006raman}. The inplane $h-$BN peak is observable in both the Raman traces indicating a proper encapsulation of the graphene layer with $h-$BN.}
\label{Figure.S7}
\end{figure*}

\begin{figure*}
\includegraphics[scale=0.7]{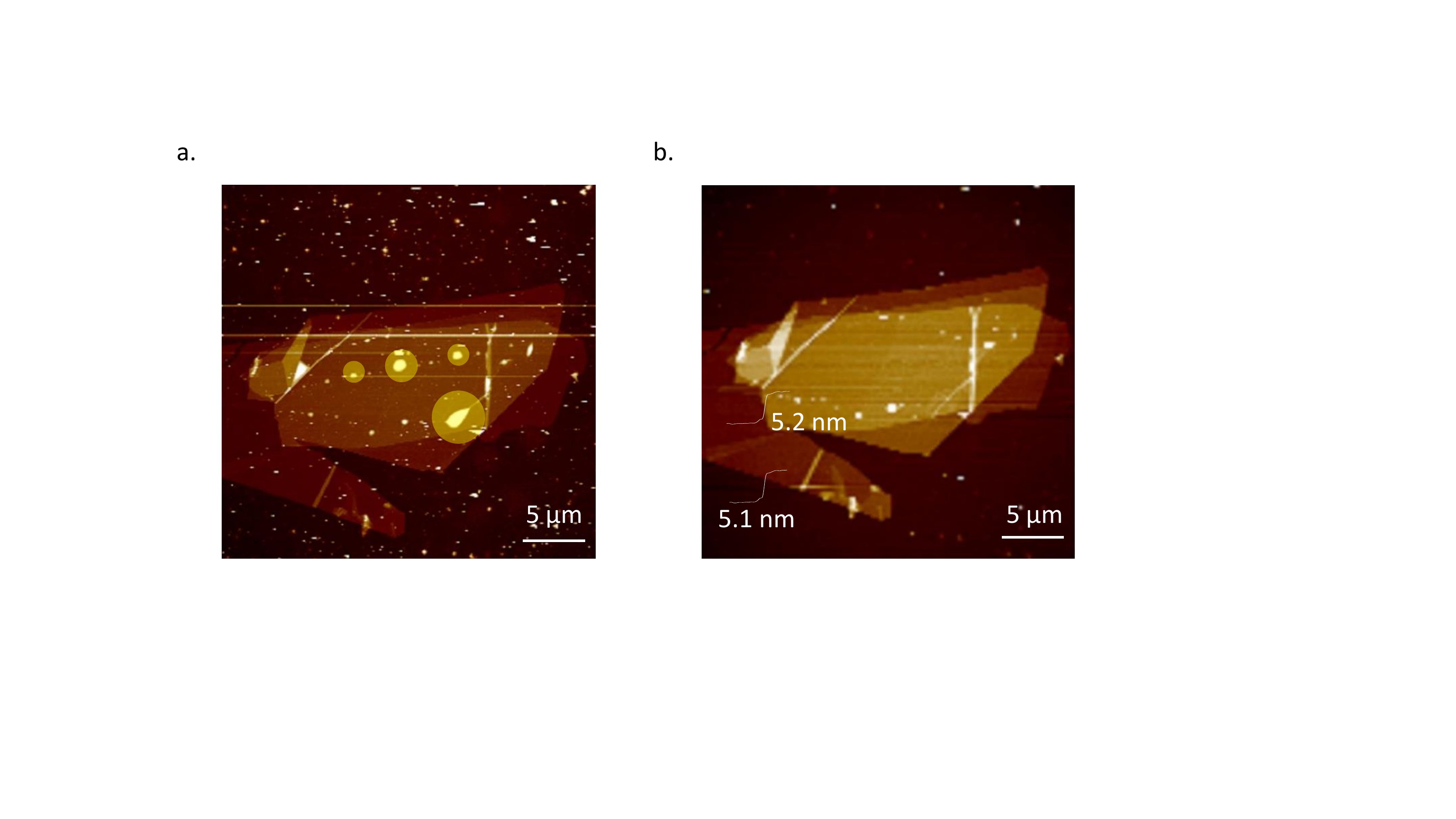}
\caption{\textbf{Atomic force microscopy (AFM) on $h$-BN encapsulated graphene heterostructure.} AFM images of the $h$-BN encapsulated graphene heterostructure in Fig.~\ref{Figure.S7} performed before \textbf{a.} and after \textbf{b.} annealing, respectively. The thickness of top and bottom $h$-BN were defined to be $\sim$5.2 nm and $\sim$5.1 nm, respectively. Comparing these two panels, we can note that annealing significantly reduces interfacial contaminants (yellow circles in \textbf{a.}) thereby improving the heterostructure quality. For the annealing process, we used H$_{\textnormal{2}}$ / Ar: 35 sccm / 200 sccm flow with 350 $^{\circ}$C for 3 hours.}
\label{Figure.S8}
\end{figure*}

\begin{figure*}
\includegraphics[scale=0.7]{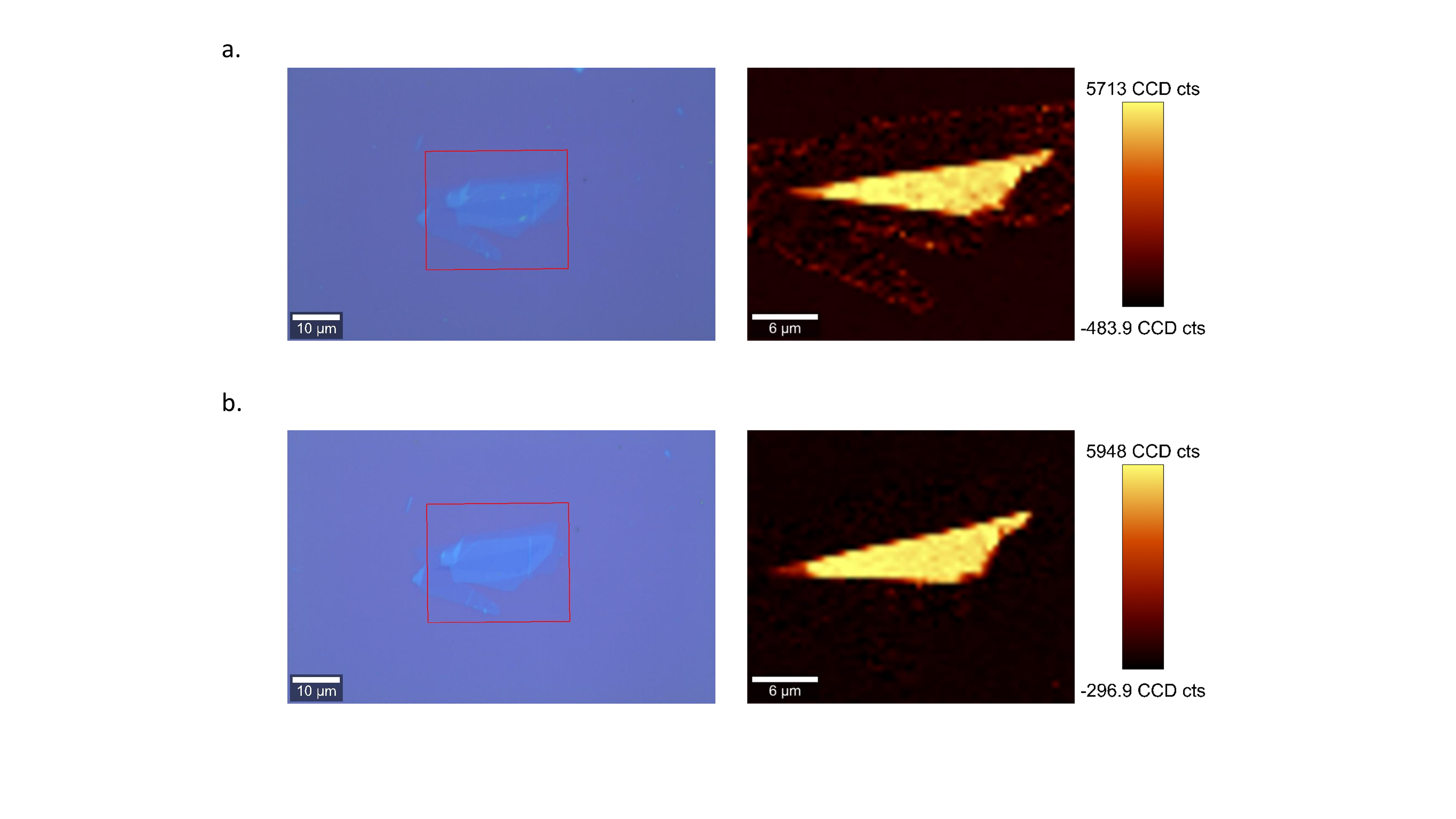}
\caption{\textbf{Raman mapping of $h$-BN encapsulated graphene heterostructure.} Raman maps performed on the $h$-BN encapsulated graphene heterostructure in Fig.~\ref{Figure.S7} before \textbf{a.} and after \textbf{b.} annealing, respectively.  Parameters of the annealing process are mentioned in Fig.~\ref{Figure.S8}. Red boxes in the left panels indicate the region of Raman mapping. The right panels map the intensity of the 2D peak which defines the graphene region. Similar Raman maps before and after annealing indicate that the crystalline quality of graphene remains unaffected by the annealing process.}
\label{Figure.S9}
\end{figure*}


\clearpage
\bibliographystyle{achemso}
\bibliography{bibliography_9}